\documentclass[aps,prl,preprint,groupedaddress,nofootinbib]{revtex4-2}
\usepackage{graphicx}
\usepackage{dcolumn}%
\usepackage{bm}
\usepackage[version=3]{mhchem}
\usepackage[left=1.5cm, right=1.5cm, top=1.785cm, bottom=2.0cm]{geometry}
\usepackage{balance}
\usepackage{amssymb}
\usepackage{mathptmx}
\usepackage{graphicx} 
\usepackage{lastpage}
\usepackage{color}
\usepackage{soul}
\usepackage{comment}
\usepackage{braket}
\usepackage[format=plain,justification=justified,singlelinecheck=false,font={stretch=1.125,small,sf},labelfont=bf,labelsep=space]{caption}
\usepackage{float}
\usepackage{fancyhdr}
\usepackage{fnpos}
\usepackage[english]{babel}
\addto{\captionsenglish}{
 
}
\usepackage{array}
\usepackage{droidsans}
\usepackage{charter}
\usepackage[T1]{fontenc}
\usepackage[usenames,dvipsnames]{xcolor}
\usepackage{setspace}
\usepackage[compact]{titlesec}
\usepackage{hyperref}
\hypersetup{
  colorlinks   = true, %Colours links instead of ugly boxes
  urlcolor     = blue, %Colour for external hyperlinks
  linkcolor    = blue, %Colour of internal links
  citecolor   = red %Colour of citations
}

\usepackage{epstopdf}

\usepackage[normalem]{ulem}
\usepackage{framed}
\makeatletter
\newcommand\footnoteref[1]{\protected@xdef\@thefnmark{\ref{#1}}\@footnotemark}
\makeatother

\newcommand{\del}[1]{\frac{\partial}{\partial #1}}

\newcommand{\kd}[1]{\delta_{#1}}
\definecolor{cream}{RGB}{222,217,201}

\begin{document}

%\title{Temporal geometry of ultrafast electronic response in chiral molecules}
\title{Geometry of chiral temporal structures I: Physical effects}

\author{Andres F. Ordonez$^{1,2}$, Aycke Roos$^{3}$, Pablo M. Maier$^{3}$, Piero Decleva$^{4,5}$ David Ayuso$^{1,3,6,7}$ and Olga Smirnova$^{3,8,9}$}

\affiliation{
%$^1$Max-Born-Institut, Max-Born-Str. 2A, 12489 Berlin, Germany\\
%$^2$ICFO-Institut de Ciencies Fotoniques, The Barcelona Institute of Science and Technology, 08860 Castelldefels (Barcelona), Spain\\
$^1$Department of Physics, Imperial College London, SW7 2BW London, United Kingdom\\
$^2$Department of Physics, Freie Universit\"at Berlin, 14195 Berlin, Germany\\
$^3$Max-Born-Institut, Max-Born-Str. 2A, 12489 Berlin, Germany\\
$^4$Dipartimento di Scienze Chimiche e Farmaceutiche, Universit\`a degli Studi di Trieste, Trieste, Italy\\
$^5$CNR-IOM DEMOCRITOS, Trieste, Italy\\
$^6$Department of Chemistry, Queen Mary University of London, E1 4NS London, United Kingdom\\
$^7$Department of Chemistry, Molecular Sciences Research Hub, Imperial College London, W12 0BZ London, United Kingdom\\
$^8$Technische Universit\"at Berlin, Straße des 17. Juni 135, 10623 Berlin, Germany\\
$^9$Technion - Israel Institute of Technology, Haifa, Israel}

\date{\today}
%\begin{comment}
\begin{abstract}
%\emph{Temporal geometry} emerges due to an opportunity to introduce
%\emph{Local} or \emph{temporal} chirality is a property of a time-dependent vector  tracing a chiral three-dimensional shape in space. % as it evolves in time. %A time-dependent vector that traces a chiral three-dimensional shape in space as it evolves in time is \emph{locally} or \emph{temporally} chiral.
In non-relativistic physics, the concepts of geometry and topology are usually applied to characterize spatial structures or structures in momentum space.
We introduce the concept of temporal geometry, which encompasses the geometric and topological properties of temporal shapes, i.e. trajectories traced by the tip of a time-dependent vector in vector space.  We apply it to vectors describing ultrafast electron currents or induced polarization in chiral molecules. %We show how one can take advantage of 
%Such temporal shapes can be understood as elements of temporal geometry. created by ultrafast electron currents in chiral molecules. %to induce, control and quantify non-linear chiral or topological response in gas phase molecules.  
The central concepts of temporal geometry  -- Berry curvature and Berry connection --  %The  curvature and  phase are well-known features of \textit{nuclear} dynamics at conical intersections in molecules, irrespective of molecular symmetry. 
emerge  as ubiquitous features of photoexcited, non-equilibrium, chiral \textit{electron} dynamics. %, even far away from conical intersections, when such  dynamics are excited 
%excited by light fields. % carrying spin angular momentum.
We demonstrate that the Berry curvature and Berry connection (i) rely on the polarization properties of light pulses, (ii) can be introduced for multiphoton processes, and (iii) control enantio-sensitive geometric observables via non-equilibrium electronic dynamics excited by tailored laser fields.
Our findings may open a way to ultrafast, topologically non-trivial, and enantio-sensitive chemical dynamics. 
%We have recently shown that geometric magnetism manifests itself in new anomalous enantio-sensitive observables in photoionization of  chiral molecules \cite{ordonez2023geometric}.  %These observables are unique messengers of electronically driven charge directed reactivity. The vectorial observables  such as the enantio-sensitive molecular orientation by ionization (PI-MOCD) are proportional to the  curvature. The scalar observables such as the enantio-sensitive component of the photoionization rate  are proportional  to the  phase. 
%Here we show that this phenomenon has broader scope and is a ubiqutous feature of non-equilibrium electron dynamics of chiral molecules. The  curvature and the  phase also emerge in bound state dynamics.  photoexcitation of chiral molecules and can be probed via circular dichroism in photoionization-induced molecular orientation.  We establish the link between the geometric magnetism in bound states and charge directed reactivity, which opens the way to ultrafast topologically non-trivial and enantio-sensitve chemical dynamics.%at conical intersections in chiral molecules.
\end{abstract}
\maketitle
\section{1. Introduction}
Synthetic chiral light \cite{Ayuso2019NatPhot} is an example of a locally chiral object: the Lissajous figure of its electric field vector at every point in space draws a chiral three-dimensional trajectory in time which cannot be superimposed onto its mirror image. Since its chirality is local, it can induce chiral interactions with electrons in molecules in the electric dipole approximation \cite{ayuso2022ultrafast}, promising enhancement of the chiral response by orders of magnitude compared to standard methods  based on inefficient interactions of electrons with the magnetic field component of light. Three-color electromagnetic fields with mutually orthogonal polarizations introduced in the pioneering work by Kral and Shapiro \cite{Kral:2001aa} and  used to harvest the electric-dipole chiral response in the microwave region \cite{Eibenberger2017PRL,Perez2017,Lee2024} also belong to this family. 

Locally chiral temporal structures do not have to be limited to the one traced by the tip of the light polarization vector. Electronic polarization or electronic currents  induced in chiral molecules \cite{Beaulieu2018PXCD,Ayuso2022OptExp} can also trace temporal chiral structures, even if the driving field is not chiral. That is, the respective vectors  -- current or induced polarization -- trace a chiral trajectory in vector space as these vectors evolve in time. The chiral shape outlined by this trajectory is local because it is encoded in the temporal evolution of the vector, rather than in its spatial evolution. The specific chiral shape is defined by the mutual orientations of the molecule and light's polarization vector: each orientation will give rise to a different temporal chiral shape. Thus, the temporally chiral shape is defined in the configuration space of molecular orientations.  If several orientations contribute to the measured response, e.g. photoionization from the current carrying superposition of states, then the set of different temporal shapes corresponding to the underlying orientations may also combine into a non-trivial global structure.   

%The possibility to realize a locally chiral response in chiral molecules poses fundamental questions about its global structure in the configuration space,  such as e.g. molecular orientations.  %Molecular orinetations which can be viewed as a global landscape. 
This poses several pertinent questions: (i) how is the temporal chiral geometry connected for two different orientations, %between two different points of a global configuration space, 
and (ii) how does this connection quantify the global response of a full set of orientations? Here, we address these questions  by identifying the Berry connection and Berry curvature associated with the \emph{temporal geometry} of the attosecond electronic response of chiral molecules to light.
%The far-field response will then encode the handedness of a molecular medium in the emission angle. For example, molecular handedness can be mapped onto the emission direction of  free induction decay \cite{Khokhlova_Sci_Adv_2022} or  harmonic generation\cite{Ayuso2021NatComm}. 
%Since different frequencies of the multicolor light combine at different points in space with different phases, such global chirality is not  guaranteed \emph{apriory}. For example, the  three colour locally chiral electromagnetic field proposed in \cite{Kral:2001aa} is globally chiral only within half of its wavelength\cite{ayuso2022ultrafast}. This is sufficient in the microwave region\cite{Patterson2013Nat}.   While various set-ups can be constructed to create locally and globally chiral light \cite{Ayuso2019NatPhot, Ordonez2023ArXiv, Ye:23}, the spatial variation of the relative phases of non-collinear multicolor fields constituting locally chiral light can also be used to shape the light's handedness in space in the near field \cite{Ayuso2021NatComm, Khokhlova_Sci_Adv_2022}.  The far-field response will then encode the handedness of a molecular medium in the emission angle. For example, molecular handedness can be mapped onto the emission direction of  free induction decay \cite{Khokhlova_Sci_Adv_2022} or  harmonic generation\cite{Ayuso2021NatComm}.
The realization of this program requires the implementation of geometric/topological concepts into the ultrafast electronic response in chiral molecules.

While topological aspects of nuclear dynamics at conical intersections in molecules are well-known \cite{mead1992geometric}, 
the connection between topology and chirality in the light-induced electronic %or optical 
response of gas-phase chiral molecules %to light 
is an emergent topic \cite{ordonez_propensity_2019-2,schwennicke2022enantioselective, peter2023chirality, ordonez2023geometric, mayer2023chiral}. In these pioneering works, the respective configuration or parameter spaces range from inverse lattice vectors (in case of periodic chiral arrangement of atoms)\cite{peter2023chirality} to parameters of electromagnetic waves triggering the response \cite{schwennicke2022enantioselective} to propagation vectors of non-linear response \cite{mayer2023chiral}.  %Geometric concepts such as curvature, connection and geometric phase often underlie new properties of matter.  
%New properties of matter frequently emerge from geometric concepts such as curvature, connection, and geometric phase.
The desire to invoke geometric concepts  such as Berry curvature, Berry connection, and geometric phase is hardly surprising as it often enables one to identify novel properties of matter. Examples range from the emergence of the Fermi anti-symmetrization principle from the topology of configuration space \cite{leinaas1977theory}, to  topological materials \cite{kane2005quantum, bernevig2006quantum}, anyons and topological quantum computing \cite{nayak2008non}, twisted light \cite{cisowski2022colloquium}, topological photonics and topological lasers \cite{RevModPhys.91.015006}.

The topological aspects of the \textit{electronic} response have so far been mainly harvested in condensed matter systems %or photonic structures
and have yet to find their room in the gas and liquid phases. 
Since topology has an attractive property of robustness to external perturbations and noise, one may be able to create efficient and robust new enantio-sensitive observables by combining chirality and topology in the ultrafast electronic and optical response of chiral gases and liquids. Relevant experimental imperfections in gas-phase experiments involve fluctuations of laser parameters, sensitivity and instrument response of photoelectron detectors, challenges in sample preparation, dilute samples and small or fluctuating enantiomeric excess in a sample. In the combination of efficiency and robustness that we are aiming to achieve, the efficiency comes from driving an ultrafast non-linear response enabling chiral detection via electric-dipole interactions \cite{ayuso2022ultrafast}, whereas robustness comes from topological concepts represented locally via Berry curvature and Berry connection in relevant configuration or parameter spaces.

We have recently identified a manifestation of the Berry connection and Berry curvature in the one-photon ionization  of chiral molecules \cite{ordonez2023geometric}. Here we extend these concepts to two-photon ionization  and focus on opportunities to control the Berry curvature and observables that it induces by controlling light field polarization.    %that the curvature in the space of laser polarization vectors has additional terms      Surprisingly, it enables a new class of extremely efficient enantio-sensitive observables \cite{ordonez2023geometric} which rely on exciting electronic or vibronic currents in chiral molecules, and serve as 
We identify and describe  enantio-sensitive observables enabled by these generalizations. As a corollary of our approach, we
establish the geometric origin of the photo-excitation circular dichroism (PXCD)\cite{Beaulieu2018PXCD}  and its detection
via photoionization with linearly polarized light (PXECD)\cite{Beaulieu2018PXCD} .
%Here, we develop an approach which is not restricted to adiabatic evolutions of molecular degrees of freedom and  geometric effects in \textit{bound} states of chiral molecules,  providing a basis for establishing geometric observables in  photoexcited non-equilibrium chiral dynamics.  As a first step in this direction, we establish the geometric origin of  photo-excitation circular dichroism (PXCD)\cite{Beaulieu2018PXCD} and its detection using photoionization by linearly polarized light (PXECD)\cite{Beaulieu2018PXCD}.
%Comment from Ling: What does this mean, concretely?  A concrete functional dependence of the response on each of these quantities?  That understanding of these quantities leads to a deeply illuminating understanding of the response?  That those quantities can be used as effective means of detection of molecular chirality?  Or something else? 
Our findings establish geometrical concepts, such as the Berry curvature and the geometric phase, as the key quantities underlying the temporally chiral electronic response. It means that (i) there is a well-defined  functional dependence of the response on these quantities, (ii) the understanding of these quantities leads to a deeply illuminating understanding of the response, (iii) those quantities can be used as effective means of detection of molecular chirality. This may open a way to generate a topologically non-trivial electronic response in photoexcited bound states of chiral molecules via tailored light fields. 

The paper is organized as follows. In \emph{Section 2} we outline the concept of \emph{temporal geometry} in the interaction of molecules with laser pulses and present  expressions for the Berry connection, Berry curvature and vectorial and scalar observables relying on the Berry curvature. % containing spin angular momentum. % and thus able to  define an oriented plane in space during the cycle of their temporal evolution. 
 In \emph{Section 3} we use  these expressions  to provide a new general (i.e. using the Berry connection and Berry curvature  from the outset) derivation of the geometric effect -- molecular orientation by photoionization (PI-MOCD)\cite{ordonez2023geometric}. We show that the direction of the Berry curvature defines the direction of molecular orientation and is a direct experimental observable. %In this effect, left (right) molecular ions orient themselves along (or opposite to) the direction of the curvature. %This general derivation allows one to see how the respective geometric contribution emerges in  any specific observable, which can be used  to probe the geometric magnetism. 
This approach shows how geometric quantities may appear in multiphoton processes and verifies our new method by comparison with the earlier results \cite{ordonez2023geometric}. It also establishes the path to derive our new results identifying the Berry curvature in bound states and in "mixed" bound and continuum states in \emph{Section 4}. %Here we apply our general method   to a specific excitation with a circularly polarized pulse. % and its probe by photoionization by \emph{linearly} polarized pulse. 
Last but not least, we show how the efficient enantio-sensitive observables generated by \emph{temporal geometry}, such as enantio-sensitive molecular orientation,  can be controlled via excitation of non-equilibrium dynamics. In \emph{Sections 2, 3, 6}, using the chiral molecule propylene oxide as an example, we demonstrate  that simple polarization control of laser pulses allows one to switch the direction of the Berry curvature from its bound to its continuum value and thus switch the direction of enantio-sensitive orientation in PI-MOCD. Both the demonstration of the control and its fundamental understanding are new.
Superscripts $M$ and $L$ denote that respective vectors belong to molecular or laboratory frame, respectively. Vectors contributing to scalar products will not be marked, because the pair of such vectors can be associated with any frame.

\section{2. Temporal geometry in chiral molecules}

\begin{figure}
\includegraphics[width=1\textwidth]{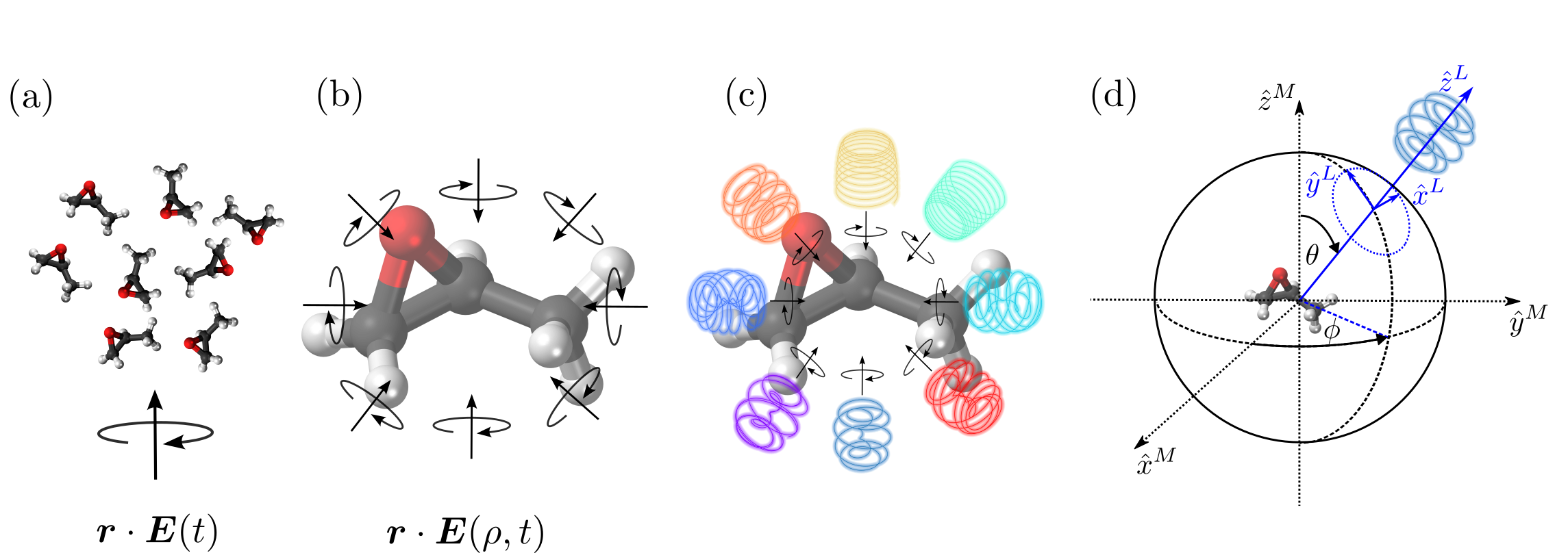}
\caption{The idea of temporal geometry. Panel (a) shows a circularly polarized field incident on an ensemble of randomly oriented molecules in the laboratory frame. In the molecular frame (panel (b)), the molecule is fixed, so it is the light that changes its orientation. Depending on the orientation, different chiral electronic currents can be induced. They are represented by glowing curves in panel (c). Panel (d): The molecular frame is defined by axes: $\hat{\bm{x}}^{M}$ $\hat{\bm{y}}^{M}$, $\hat{\bm{z}}^{M}$. The laboratory frame is defined by the polarization plane $\hat{\bm{x}}^{L}$ $\hat{\bm{y}}^{L}$ and the light propagation vector $\hat{\bm{z}}^{L}$. The polarization plane is tangent to the sphere at the points ($\theta,\phi$). The third angle $\alpha$ (not shown) defines the orientation of the polarization ellipse, which is not relevant when considering a monochromatic circularly polarized field. The blue curve represents a chiral 3D shape drawn by the electronic current excited by the laser field with polarization on the plane tangent to the sphere in points ($\theta,\phi$).  }
\label{fig:1}
\end{figure}

Light fields $\bm{E}(t)$ interacting with molecules primarily couple to electrons and trigger their attosecond response. In the laboratory frame\footnote{To define the laboratory frame in the electric-dipole approximation, we cannot rely on the propagation vector of the laser field. Instead, we can define the laboratory frame using the two orthogonal polarization components of the laser field together with their cross product, necessitating circularly or elliptically polarized fields. \label{footnote:a}} circularly polarized light interacts with randomly oriented molecules (Fig.\,\ref{fig:1}(a)). Since the strength of such interaction depends on the orientation of a molecule with respect to the laser field, it inevitably couples electronic and rotational degrees of freedom. It means that the electronic current excited in  molecules oriented differently will be different.  It is easier to visualize it by switching to the molecular frame, in which the laser field would have all possible orientations (Fig.\,\ref{fig:1}(b)). The coupling term explicitly depends on the Euler angles, which characterize the orientation of the laser field in the molecular frame. %This is our configuration space.
% in molecules.
%The the rotational dynamics is described 
%\begin{equation}
 %i\frac{\partial}{\partial t}\psi_{rot}(t,\rho)=H_{rot}\psi_{rot}(t,\rho).
 %\label{eq:TDSE_rot}
%\end{equation}
%For example, consider interaction of light with a randomly oriented molecule. 
%In the molecular frame light's  polarization vector changes its orientation accordingly.  For example, 

To illustrate the geometric origin of such coupling, it is convenient to associate  the normal to the light polarization plane in the molecular frame with a unit radius vector on a sphere characterized by angles ($\theta, \phi$).  The polarization plane is tangent to this sphere. Therefore, the light's polarization vector 
in the molecular frame $\bm{e}({\theta, \phi, \alpha})$ depends on three angles, where %angle 
 %$\theta, \phi$ characterize the orientation of propagation vector on the sphere, while angle
$\alpha$ describes the orientation of the polarization ellipse \footnote{In case of a circularly polarized field the angle $\alpha$ is redundant. We omit $\alpha$ in the following.} in the tangent plane (Fig.\,1).  %Thus the interaction term  in the Hamiltonian in molecular frame parametrically depends on  
Any change in the molecular orientation will "transport" the polarization vector along the surface of the sphere, while the electronic wave-function describing excited electrons will depend on this transport. At any point on the sphere, the laser field excites a geometrically different \emph{temporally chiral} electronic response, %analogous to a signal detected in Ref.\,\cite{Beaulieu2018PXCD},
because the geometry of light polarization vectors as seen by the molecule (Fig.\,1(b,c)) changes from one point on the sphere to another. Is there a global geometric property associated with these chiral currents for different molecular orientations?

%Formally, the  light field is a tangent field on the space (manifold) of molecular orientations, represented by the surface of the sphere. Mathematically, a connection is a differential operator $\hat{A}=\bm{\nabla}_{\bm{E}}$ defined on vector valued functions, such as $\bm{E}(t,{\theta, \phi, \alpha})$,which is used to define such transport by connecting different geometries, excited in different $(\theta, \phi)$ points.

%Question from Ling: Is this a single electron wave function?  If yes, the electron is the one that was excited by the light field? 

%I am also trying to reconcile this wave function picture with the picture of "chiral current."  The former gives me a picture of a time-dependent distribution of the electron density.  So, in principle I would use the concept of local current density based on this picture.  The latter evokes a sense of a single particle moving in space.  How to connect these two?

Solving the time-dependent 
 Schrödinger equation (TDSE) locally for each  orientation characterized by a point $\rho_0 = (\theta_0, \phi_0)$ $\textcolor{blue}{^3}$ on the sphere of Euler angles in the electric dipole approximation:  
\begin{equation}\label{eq:TDSE_local}
i\frac{d}{dt}\psi_{el}^{0}(\bm{r},\rho_0,t)=\bigg[H_{el}+\bm{r}\cdot\bm{E}(\rho_0,t)\bigg]\psi_{el}^{0}(\bm{r},\rho_0,t),
\end{equation} 
we obtain an electron   
 wave-function $\psi_{el}^{0}(t,\bm{e}(\rho_0),\bm{r})$ \footnote{For simplicity we consider a single active electron problem.}, which is "agnostic" about other light field geometries.
 The temporally chiral shapes are generated by the induced polarization $\bm{P}(\rho,t)$ or current $\bm{j}(\rho,t)=-\frac{\partial \bm{P}}{\partial t}$ vectors: $\bm{P}(\rho,t)\equiv -\langle \psi_{el}^{0}(t,\bm{e}(\rho),\bm{r})|\hat{r}|\psi_{el}^{0}(t,\bm{e}(\rho),\bm{r})\rangle$.
Thus, the  question about the connection between the two currents at two different orientations reduces to the question about the connection between the two wave functions for two different $\rho$. Can these "local" solutions be connected in any way? Is there a Berry curvature in the configuration space of molecular orientations, e.g. for randomly oriented molecules?

The comparison between the electron wave-function $\psi_{el}^{0}(t,\bm{e}(\rho),\bm{r})$  describing the temporal geometry of chiral currents excited by the laser field fixed at a given point on the sphere $\rho_0 = (\theta_0, \phi_0)$  with its "global" counterpart $\psi_{el}(t,\bm{e}(\rho),\bm{r})$, "informed" about all different laser geometries,  encapsulates the essence of the "connection" between different local geometries. 
%Comment from Ling: This entire sentence is too long and contains too many pieces of information.  It is very difficult to read.  I suggest that it be broken into a few sentences.  Some provide clear definitions of the involved quantity.  Other(s) explain other information such as reasoning or significant knowledge that readers are being given.
The TDSE for the electron wave-function $\psi_{el}^{0}(t,\bm{e}(\rho_1),\bm{r})$ at some other point on the sphere has the same general form, but the laser field, of course, has a different geometry leading to a different temporally chiral response. The issue arises due to the fact that these two TDSEs are completely independent, and thus each "local" electron wave-function $\psi_{el}^{0}(t,\bm{e},\bm{r})$ is defined up to an arbitrary phase. It means that one may need to find a consistent procedure to "connect" all such solutions along a given cyclic path in the configuration space by introducing a common phase connecting all local solutions in a well-defined way and encoded in a single, consistent "full" solution $\psi_{\rm el}(t)$:
%\begin{equation}
%\Psi_{el}(\bm{r},T,t)=e^{iS(T)}\psi_{el}^{0}(t,\bm{E}%(\overline{\rho}(T),t),\bm{r}).
% \label{eq:substitution}
%\end{equation}
\begin{align}\label{eq: dynlift}
    \ket{\psi_{\rm el}(t)} &= e^{-i\alpha_{\rm D}(t) + iS(t)} \ket{\psi_{\rm el}^0(t)}\,.  
    %\label{eq:substitution}
\end{align} 
In general, the phase has a dynamical $\alpha_{\rm D}(t)$
\begin{align} 
    \alpha_{\rm D}(t) &= \int\limits_0^t d\tau \bra{\psi_{\rm el}(\tau)}H(\tau)\ket{\psi_{\rm el}(\tau)}\,,
\end{align}
and a geometric  $S(t)$
\begin{align} 
    S(t) &= i\int\limits_0^t d\tau \bra{\psi_{\rm el}^0(\tau)}\del{\tau} \ket{\psi_{\rm el}^0(\tau)}\, 
\end{align}
components. Our "local" solution $\psi_{\rm el}^0(t)$ is known as a \emph{closed lift} of a state vector defined as $\ket{\psi_{\rm el}(t)}\bra{\psi_{\rm el}(t)}$ (see \cite{aharonov_anandan} and Appendix A). In Appendix B we show that, in the case of a periodic evolution of the polarization vector $\bm{e}({\rho})$, the Berry connection is a vector potential 
\begin{equation}
 \bm A(\bm{e})=i\langle \psi_{el}^{0}(\bm{e})|\bm{\nabla}_{\bm{e}}\psi_{el}^{0}(\bm{e})\rangle,
 \label{eq:connection_t}
\end{equation}
quantifying the change in the local wave-function in response to the change in the laser field geometry. 
It specifies the geometric phase $S$ accumulated after a full cycle of evolution of $\bm{e}({\rho})$ in the space of Euler angles $\rho$ as detailed in Appendix B:
\begin{equation}
 S=\oint i\langle \psi_{\rm el}^0(\bm{e})|\bm{\nabla}_{\bm{e}}\psi_{\rm el}^0(\bm{e})\rangle d\bm{e}=\oint \bm A(\bm{e})\cdot d\bm{e}.
 \label{eq:S}
\end{equation}
%and
%\begin{equation}
%\Psi_{el}(\bm{r},T,t)\overset{def}=\int d\rho |\psi_{rot}%(\rho,T)|^2\psi_{el}^{0}(\bm{r},\rho,t).
%\label{eq:global_el}
%\end{equation}
%Here $\psi_{rot}(\rho,T)$ describes a generic (not restricted to the adiabatic limit) evolution of rotational wave-function driven by an arbitrary Hamiltonian $H_{rot}(t)$,    $\overline{\rho}(T)=\int d\rho |\psi_{rot}(\rho,T)|^2\rho$ and $\psi_{el}(\bm{r},\rho,t)$ solves the "local" TDSE for the electronic degree of freedom:
%\begin{equation}\label{eq:TDSE_local}
%i\frac{d}{dt}\psi_{el}^{0}%(\bm{r},\rho,t)=\bigg[H_{el}+\bm{r}\cdot\bm{E}(\rho,t)\bigg]\psi_{el}^{0}(\bm{r},\rho,t).
%\end{equation}
%\par
%Thus, once applied to electron’s "local" wave-functions, the connection takes a shape of a vector-potential
Eq.\,(\ref{eq:connection_t}) encapsulates the origin of geometric magnetism, emerging due to the necessity to "connect" \emph{temporal  geometry} at different points in configuration space. The gradient with respect to the light polarization vector in Eq.\,(\ref{eq:connection_t}) emphasizes the dynamical, laser-driven origin of these geometric quantity. As derived, Eq.\,(\ref{eq:connection_t},\ref{eq:curvature_def}) appear applicable to any light-driven process. 

Due to the complex nature of the polarization vector the Berry curvature takes the form of a skew-hermitian tensor $\Omega^{ij}= i \braket{\partial_{e_i}\psi_{\mathrm{el}}^0(\bm{e})|\partial_{e_j}\psi_{\mathrm{el}}^0(\bm{e})}$. Naturally, it can be partitioned into a real valued, antisymmetric component $\Omega_a^{ij}$ and a imaginary valued, symmetric component $\Omega_s^{ij}$, respectively. Evaluating the curl of the Berry connection we obtain an antisymmetric part of the Berry curvature tensor (see companion paper \cite{roos2025geometrychiraltemporalstructures}):
\begin{equation}
 \bm \Omega=\bm{\nabla}_{\bm{e}}\times \bm{A}= i\langle \bm{\nabla}_{\bm{e}}\psi_{\mathrm{el}}^{0}(\bm{e})|\times |\bm{\nabla}_{\bm{e}}\psi_{\mathrm{el}}^{0}(\bm{e})\rangle, 
\label{eq:curvature_def}
\end{equation}
such that $\Omega_a^{ij} = \frac{i}{2}\varepsilon^{lij}\bm{\Omega}_l$. Below, we use Eq.\,(\ref{eq:curvature_def}) to quantify the ability to control molecular orientation in two photon excitation for different configurations of pump and probe pulse polarizations.  First, we derive the relationship between the orbital antisymmetric Berry curvature (Eq.\,(\ref{eq:curvature_def})) and the vectorial  observables expressing the 
light-induced enantio-sensitive molecular orientation by ionization of an initially randomly oriented ensemble of molecules, generalizing  our earlier result \cite{ordonez2023geometric}, where we introduced the orbital antisymmetric Berry curvature for one-photon ionization from a state with a stationary current. Second, we identify new  regimes  for observing   enantio-sensitive orientation in two-photon ionization, where selection of preferred orientation happens at the excitation step, opening new  opportunities for controlling the direction of orientation. %Third, we derive the relationship between the photoionization yield and the curvature Eq.\,(\ref{eq:curvature_def}) in two-photon ionization by circularly polarized field. We show that it is similar to the one in one-photon ionization from a state with a stationary current \cite{ordonez2023geometric}.%, confirming the generality of the link between the enantio-sensitive photoionization yield and the geometric phase. 

Specifically, we show that enantio-sensitive molecular orientation can be expressed using  Eq.\,(\ref{eq:curvature_def}) as follows
(superscripts $L$ and $M$ stand for laboratory and molecular frame correspondingly, and we consider a driving field polarised on the $x^M-y^M$-plane):
%As we demonstrate in Sections 3-6 below, the curvature leads to new vectorial observables expressing the light induced enantio-sensitive molecular orientation by ionization or by excitation of an initially randomly oriented ensemble of molecules (superscripts $L$ and $M$ stand for laboratory and molecular frame correspondingly):
\begin{equation} \left\langle\bm{v}_{\Omega}^{L}\right\rangle\propto \sigma\left\langle\left(\bm \Omega^L\cdot\hat{\bm{z}}^{L}\right)\bm{v}_{\Omega}^{L}\right\rangle=\sigma\mathrm{R}^{(n,p)}\langle \bm{\Omega}^M\cdot\bm{v}_{\Omega}^M\rangle\hat{\bm{z}}^{L}.
\label{eq:observables_intro}
\end{equation}
Here $\left\langle\bm{v}_{\Omega}^{L}\right\rangle$ is the orientationally averaged value of a unit vector directed along the Berry curvature in the molecular frame ($\bm{v}_{\Omega}^{M}\parallel\bm{\Omega}^M$), $\sigma=\pm1$ is the direction of rotation of the light field,  %We also show that degree of orientation Eq.\ref{eq:observables_intro} can be expressed in a simpler form emphasising its 
%These expressions emphasise the origin of enantio-sensitive molecular orientation linked to the existence of the curvature.  The second expression quantifies the direction and degree of molecular orientation. 
%\begin{equation} \left\langle\bm{v}_{\Omega}^{L}\right\rangle=\sigma\mathrm{R}^{(n,p)}\langle \bm{\Omega}^M\cdot\bm{v}_{\Omega}^M\rangle\hat{\bm{z}}^{L}.
%\label{eq:observables_intro}
%\end{equation}
%\begin{equation} \left\langle\bm{v}_{\Omega}^{L}\right\rangle=\sigma \left\langle\left(\bm \Omega\cdot\hat{\bm{z}}^{L}\right)\bm{v}_{\Omega}^{L}\right\rangle=\sigma\mathrm{R}\left(\langle\bm \Omega\rangle\cdot\bm{v}_{\Omega}^{M}\right)\hat{\bm{z}}^{L}=\sigma\mathrm{R}\upsilon|\langle\bm \Omega\rangle|\hat{\bm{z}}^{L}.
%\label{eq:observables_intro}
%\end{equation}
 %$\upsilon=\pm1$ is molecular pseudoscalar $\bm{\Omega}\cdot\bm{v}_{\Omega}\equiv \upsilon|\bm{\Omega}|$, $\langle|\bm \Omega|\rangle$ is the orientationally averaged absolute value of the curvature $|\bm \Omega|$, 
$\langle\bm{\Omega}^M\cdot\bm{v}_{\Omega}^M\rangle$ is the orientationally averaged value of the scalar product $\bm{\Omega}^M\cdot\bm{v}_{\Omega}^M$,
$\mathrm{R}^{(n,p)}$ is a factor associated with the partial alignment of the molecular ensemble due to excitation or ionization, which depends on the number of absorbed photons $n$. The superscript $p$ in $\mathrm{R}^{(n,p)}$ indicates that it also depends on the polarizations of the absorbed photons, which may or may not break the cylindrical symmetry of the set-up.  We establish the factor $\mathrm{R}^{(n,p)}$ by explicitly evaluating $\left\langle\left(\bm \Omega^L\cdot\hat{\bm{z}}^{L}\right)\bm{v}_{\Omega}^{L}\right\rangle$ and $\langle \bm{\Omega}^M\cdot\bm{v}_{\Omega}^M\rangle$ using Eq.\,\eqref{eq:curvature_def}. Importantly, factor $\mathrm{R}^{(n,p)}$ does not affect the direction of enantio-sensitive orientation unless the excitation step strongly breaks the isotropy of the initial molecular ensemble: in this work we only capture isotropic effects. In this case,Eq.\,\eqref{eq:observables_intro} shows that (i) the vector $\left\langle\bm{v}_{\Omega}^{L}\right\rangle$ orients itself along the direction of photon spin $\hat{\bm{z}}^{L}$, and that (ii)  the orbital antisymmetric Berry curvature controls the direction and the degree of molecular orientation. %The first equality in Eq.\,\ref{eq:observables_intro} is general and applicable to any multiphoton process. 
We use Eqs.\,(\ref{eq:curvature_def},\ref{eq:observables_intro}) to evaluate the direction of enantio-sensitive molecular orientation for a range of two-photon processes and compare the resulting expressions with the direct evaluation, where the orbital antisymmetric Berry curvature is not employed.  We show that the results are equivalent. Our approach not only allows us to express enantio-sensitive molecular orientation in a compact way (Eq.\,\eqref{eq:observables_intro}), but also uncovers its fundamental origin, appearing to be invariant for different  multiphoton processes.

\section{3. Temporal geometry in Two-Photon Ionization}
We can use Eq.\,\eqref{eq:curvature_def} to calculate the Berry curvature for a two-photon process, in which a linearly polarized pulse excites a superposition of states in a molecule and a circularly polarized pulse photoionizes this superposition.  Since only spin carrying light fields induce temporal geometry, we use a linearly polarized pulse to excite the molecule to separate the Berry curvature due to the continuum states from the Berry curvature due to the bound states (see Section 4). If excitation and ionization are performed with the same circularly polarized pulse, then the contributions of the bound and continuum states to the Berry curvature are mixed (see Section 5).

We  use perturbation theory to calculate $|\psi(\bm{E})\rangle$ explicitly: 
 \begin{align}
 \label{eq:psi}
\psi & =\psi_{0}+a_{1}\psi_{1}+a_{2}\psi_{2}+\int\mathrm{d}\Theta_{k}a_{\bm{k}}\psi_{\bm{k}}.
\end{align}
Here $\mathrm{d}\Theta_{k}=\sin\theta_{k}\mathrm{d}\theta_k\mathrm{d}\phi_k$ and $\theta_{k}$, $\phi_{k}$ are angles characterizing the direction of the photoelectron momentum $\bm{k}$ for fixed $k=|\bm{k}|$, $a_i$ and $a_{\bm{k}}$ are the amplitudes of bound ($i=1,2$)  and continuum states:
%\begin{equation}
%a_{i}=A_{i}^{\left(1\right)}\left(\bm{d}_{i}\cdot\bm{\mathcal{E}}\right)
%\end{equation}
\begin{equation}
 a_{i}=i\int_0^t dt' \bm{d}_{{{i}}}\cdot \bm{\mathcal{E}}_{i}(t')e^{i\omega_{i} t'}=i\bm{d}_{i}\cdot{\bm{\mathcal{E}}}_{i},%\\
 \end{equation}
\begin{equation}
a_{\bm{k}}=-\left(\bm{D}_{1}\cdot\bm{E}_{{\bm{k}1}}\right)\left(\bm{d}_{1}\cdot\bm{\mathcal{E}}_{1}\right)e^{-i\omega_1\tau}-\left(\bm{D}_{2}\cdot\bm{E}_{{\bm{k}2}}\right)\left(\bm{d}_{2}\cdot\bm{\mathcal{E}}_{2}\right)e^{-i\omega_2\tau}.
\label{eq:a_k}
\end{equation}
%\begin{align}
%\nabla_{\bm{E}}\psi & =\int\mathrm{d}\Omega_{k}\left[A_{1}^{\left(2\right)}\left(\bm{d}_{1}\cdot\bm{\mathcal{E}}\right)\bm{D}_{1}+A_{2}^{\left(2\right)}\left(\bm{d}_{2}\cdot\bm{\mathcal{E}}\right)\bm{D}_{2}\right]\psi_{\bm{k}}\\
 %& =\int\mathrm{d}\Omega_{k}\left(\mathcal{A}_{1}\bm{D}_{1}+\mathcal{A}_{2}\bm{D}_{2}\right)\psi_{\bm{k}}\\
 %& =\int\mathrm{d}\Omega_{k}\bm{\mathcal{D}}_{\bm{k}}\psi_{\bm{k}}
%\end{align}
Here $\bm{d}_{i}$ are bound transition dipoles, and $\bm{D}_{i}$ are photoionization dipoles from bound states ($i=1,2$) to the continuum state $\bm{k}$. %add reference to companion paper (next sentence)
The only relevant (for isotropic signal, i.e. independent on the third Euler angle $\alpha$) component of the antisymmetric Berry curvature is the one in the space of circular polarization vectors $\bm{e}$~\cite{roos2025geometrychiraltemporalstructures}. Writing $\bm{E}_{{\bm{k}i}}=|E_{\omega_{\bm{k}i}}|\bm{e}$, with $\bm{e}=\frac{1}{\sqrt{2}}\left(\hat{\bm{x}}+i\sigma\hat{\bm{y}}\right)$ spanning the parameter space, the Berry curvature (Eq.\,(\ref{eq:curvature_def})) in the molecular frame yields
%\begin{align}
%\bm{\Omega}_{\mathrm{ion}}(k,\rho)%&=i\langle\bm{\nabla}_{\bm{E}}\psi|\times|\bm{\nabla}_{\bm{E}}\psi\rangle \\\nonumber
%& =-2|E_{\omega_{k1}}||E_{\omega_{k2}}|\left(\bm{d}_{1}\cdot\bm{\mathcal{E}}_{1}\right)^{*}\left(\bm{d}_{2}\cdot\bm{\mathcal{E}}_{2}\right)\bm{\mathsf{P}}^{+}_{12}(k)\sin(\omega_{12}\tau),
%\label{eq:_L_continuum}
%\end{align}
\begin{align}
\bm{\Omega}_{\mathrm{ion}}^M(k,\rho)%&=i\langle\bm{\nabla}_{\bm{E}}\psi|\times|\bm{\nabla}_{\bm{E}}\psi\rangle \\\nonumber
&=-|E_{\omega_{\bm{k}1}}||E_{\omega_{\bm{k}2}}|\mathrm{Im} \left\{\left(\bm{d}_{1}\cdot\bm{\mathcal{E}}_{1}\right)^{*}\left(\bm{d}_{2}\cdot\bm{\mathcal{E}}_{2}\right) e^{i\omega_{12}\tau}\int\mathrm{d}\Theta_{k}\bm{D}_{1}^{*M}\times\bm{D}_{2}^M\right\}
\label{eq:_L_continuum}
\end{align}
%\begin{align}
%\bm{\Omega}^{L}(k,\rho)=-\langle\bm{\nabla}_{\bm{E}}\psi|\times|\bm{\nabla}_{\bm{E}}\psi\rangle & 
%=2\emph{Im} \left[ A_{1}^{*\left(2\right)}A_{2}^{\left(2\right)}\left(\bm{d}_{1}\cdot\bm{\mathcal{E}}\right)\left(\bm{d}_{2}\cdot\bm{\mathcal{E}}\right)e^{i\omega_{12}\tau}\int\mathrm{d}\Omega_{k}\bm{D}_{1}^{*}\times\bm{D}_{2}\right]
%\end{align}
%\begin{equation}
%\mathcal{A}_{i}\equiv A_{i}^{\left(2\right)}\left(\bm{d}_{i}\cdot\bm{\mathcal{E}}\right)
%\end{equation}
%\begin{equation}
%\bm{\mathcal{D}}_{\bm{k}}\equiv\left(\mathcal{A}_{1}\bm{D}_{1}+\mathcal{A}_{2}\bm{D}_{2}\right)
%\end{equation}
where the subscript 'ion' emphasizes that the Berry curvature $\bm{\Omega}_{\mathrm{ion}}^M(k,\rho)$ relies on the properties of continuum states contributing to photoionization dipoles and thus manifests itself due to geometry of continuum currents during photoionization.  
The superscript $M$ emphasizes that vectors are expressed in the molecular frame. The direction of the Berry curvature in the molecular frame is defined by the direction of 
the vector $\bm{\mathsf{P}}^{+M}_{12}(k)$, where $\bm{\mathsf{P}}^{+M}_{12}(k)\equiv\frac{1}{2}\mathrm{Im} \left\{ i\int\mathrm{d}\Theta_{k}\bm{D}_{1}^{*M}\times\bm{D}_{2}^M\right\}$, superscript "+" indicates that only $\bm{k}$-even part of the vector product contributes to the integrated quantity (see also \cite{ordonez2023geometric}). 
\begin{figure}
\includegraphics[width=.86\textwidth, angle=0]{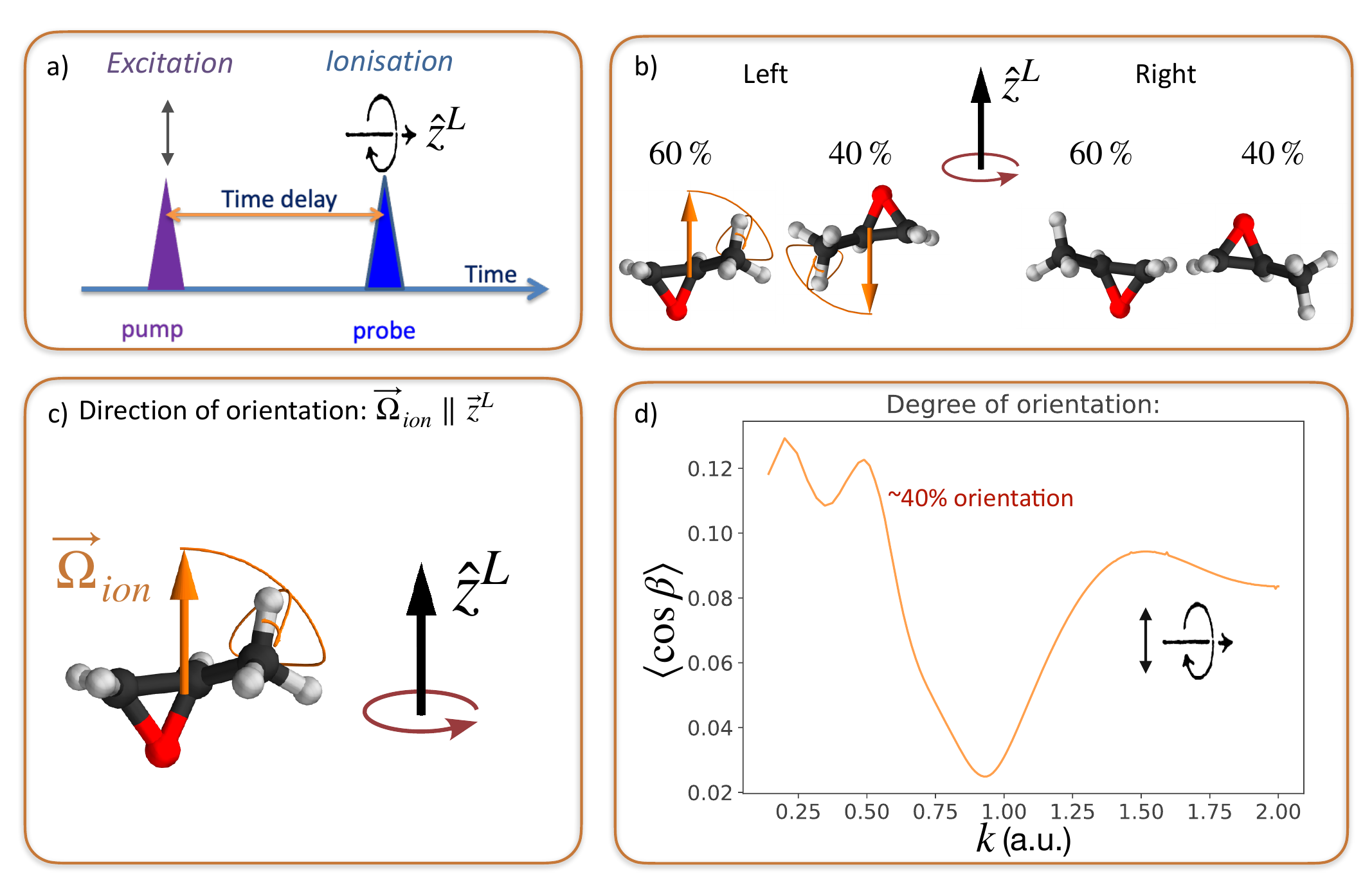}
\caption{Enantio-sensitive molecular orientation by photoionization. (a) A linearly polarized pump pulse excites the superposition of LUMO and LUMO$_{+1}$ states in propylene oxide, and then a circularly polarized probe ($\hat{\bm{z}}^L$ is the direction of photon spin in the laboratory frame) removes the excited electron. (b) Molecules with the Berry curvature $\bm{\Omega}_{ion}\parallel \hat{\bm{z}}^L$ are preferentially ionized, leading to an enantio-sensitive delay-dependent orientation of cations. $\bm{\Omega}_{ion}$ depends on the excited superposition and photoelectron energy. As a function of photoelectron energy, the tip of the Berry curvature vector traces the orange curve in space. This orange curve illustrates the sequence of orientations achieved for different final photoelectron momenta, from 0.25 to 2 a.u. Estimated  percentages of "up" and "down" oriented species correspond to the maximal value of  $\langle \cos \beta\rangle\simeq 0.11$ and obtained using the same model as in Ref.\,\cite{ordonez2023geometric}.  (c) Direction of molecular orientation is defined by the Berry curvature vector. (d) Expectation value  $\langle \cos \beta\rangle$ as a function of photoelectron momentum $k$ (in atomic units). The values have been taken at the time-delay that maximizes $\left\langle\bm{v}_{\Omega}^{L}\right\rangle$  and have been normalized to the total yield averaged over the pump-probe time delay }
\label{fig:ion}
\end{figure}

We now introduce the polar unit vector $\bm{v}_{\Omega}^M$ along the direction of the Berry curvature $\bm{\Omega}_{\mathrm{ion}}^M(k,\rho)$ (i.e. along $\bm{\mathsf{P}}^{+M}_{12}(k)$) and obtain the averaged value of $\langle\bm{\Omega}_{\mathrm{ion}}\cdot\bm{v}_{\Omega}\rangle$ \footnote{Here and in the following we omit indices $M$, $L$ in scalar products, since the scalar products are rotationally invariant and can be associated with any frame. }  (see Appendix D):
\begin{align}
\langle\bm{\Omega}_{\mathrm{ion}}(k)\cdot\bm{v}_{\Omega}\rangle%&=i\langle\bm{\nabla}_{\bm{E}}\psi|\times|\bm{\nabla}_{\bm{E}}\psi\rangle \\\nonumber
&=-\frac{1}{3}C\left(\bm{d}_{1}\cdot\bm{d}_{2}\right)\left(\bm{\mathsf{P}}^{+}_{12}(k)\cdot\bm{v}_{\Omega}\right)\sin(\omega_{12}\tau)=-\frac{1}{3}\upsilon C\left(\bm{d}_{1}\cdot\bm{d}_{2}\right)|\bm{\mathsf{P}}^{+}_{12}(k)|\sin(\omega_{12}\tau),
\label{eq:_L_continuum_av}
\end{align}
where $C\equiv|E_{\omega_{\bm{k}1}}||\mathcal{E}_{\omega_{1}}||E_{\omega_{\bm{k}2}}||\mathcal{E}_{\omega_{2}}|$.
Note that $\bm{\mathsf{P}}^{+}_{12}(k)\cdot\bm{v}_{\Omega}=\upsilon|\bm{\mathsf{P}}^{+}_{12}(k)|$, where $\upsilon=\pm1$ is a pseudoscalar characterizing the molecular handedness.
The enantio-sensitive molecular orientation by ionization (PI-MOCD) for the linear pump - circular probe excitation is associated with the preferential photoionization of the molecules with the Berry curvature  oriented along the photon spin. 
Using Eq.\,\eqref{eq:observables_intro} together with Eq.\,\eqref{eq:_L_continuum} we can calculate (see Appendix D) the orientation averaged value of $\left\langle\bm{v}_{\Omega}^{L}\right\rangle$:
 \begin{equation} \left\langle\bm{v}_{\Omega}^{L}\right\rangle^{\mathrm{\updownarrow,\circlearrowleft}}=N\sigma \left\langle\left(\bm \Omega_{\mathrm{ion}}^{L}\cdot\hat{\bm{z}}^{L}\right)\bm{v}_{\Omega}^{L}\right\rangle=N\sigma\mathrm{R}^{(2,\mathrm{\updownarrow,\circlearrowleft})}\langle\bm \Omega_{\mathrm{ion}}\cdot\bm{v}_{\Omega}\rangle\hat{\bm{z}}^{L}.
\label{eq:observables}
\end{equation}
Here $N$ is a normalization factor, $N^{-1}=\int_0^{2\pi} d\phi_{\tau}\int d\rho W^{\mathrm{\updownarrow,\circlearrowleft}}(\rho)$, where $W^{\mathrm{\updownarrow,\circlearrowleft}}(\rho)$ is photoionization rate for a given molecular orientation $\rho$, $\int_0^{2\pi} d\phi_{\tau}$ describes the averaging over delay $\tau$ between the pump and probe pulses, and 
\begin{equation}
R^{(2,\mathrm{\updownarrow,\circlearrowleft})}\equiv\frac{2}{5}\left[1 - \frac{1}{2}\frac{(\bm{d}_{1}\cdot\bm{v}_{\Omega})(\bm{d}_{2}\cdot\bm{v}_{\Omega})}{(\bm{d}_{1}\cdot\bm{d}_{2})}\right].
\label{eq:orientation_full_factor_new}	
 \end{equation} 
  Up to notations, Eq.\,\eqref{eq:observables}  reproduces our earlier result for PI-MOCD \cite{ordonez2023geometric},
validating the general approach  (i.e. calculating  $\left\langle\bm{v}_{\Omega}^{L}\right\rangle$ using Eqs.\,(\ref{eq:curvature_def} and \ref{eq:observables_intro})  developed here). For completeness we also provide the result calculated without employing the concept of the orbital antisymmetric Berry curvature, i.e. only using the definition 
 of the oriented vector $\left\langle\bm{v}_{\Omega}^{L}\right\rangle^{\mathrm{\updownarrow,\circlearrowleft}}=N\int d\rho W^{\mathrm{\updownarrow,\circlearrowleft}}(\rho)\bm{v}_{\Omega}^{L}$:
\begin{align}
\left\langle\bm{v}_{\Omega}^{L}\right\rangle^{\mathrm{\updownarrow,\circlearrowleft}} & =\frac{i\sigma}{60}NC\int\mathrm{d}\Omega_{k}\left[4\left(\bm{d}_{2}\cdot\bm{d}_{1}\right)\bm{v}_{\Omega}-\left(\bm{d}_{2}\cdot\bm{v}_{\Omega}\right)\bm{d}_{1}-\left(\bm{d}_{1}\cdot\bm{v}_{\Omega}\right)\bm{d}_{2}\right]\cdot\left(\bm{D}_{2}^{*}\times\bm{D}_{1}\right)e^{i\omega_{21}\tau}\hat{\bm{z}}+\mathrm{c.c.}\label{eq:lin-circ_exact}
\end{align}

Note that the first term in the expression for $R^{(2,\mathrm{\updownarrow,\circlearrowleft})}$ corresponds to the contribution to the signal from the the isotropic ensemble and the second term corresponds to the contribution emerging due to the alignment of molecular ensemble by the linearly polarised field. As long as $(\bm{d}_{1}\cdot\bm{d}_{2})\neq0$ the isotropic term dominates and the orientation is controlled by the orbital antisymmetric Berry curvature. The effect is maximised when $\bm{d}_{1}\parallel\bm{d}_{2}$. If $(\bm{d}_{1}\cdot\bm{d}_{2})=0$, the isotropic contribution vanishes and  the direction of enantio-sensitive orientation is controlled by vectors $\bm{d}_{1}$ and  $\bm{d}_{2}$, as it is clear from the last two terms in Eq.\eqref{eq:lin-circ_exact}.

 The vector $\left\langle\bm{v}_{\Omega}^{L}\right\rangle$ can be used to find the expectation value of $\cos\beta$, where $\beta$ is the angle between the unit vector along the Berry curvature $\bm{v}_{\Omega}^{L}$ and the direction of photon spin.  The value of $\cos\beta$ (see Fig.\,\ref{fig:ion}(c)) is given by the normalized (to the total photoionization yield averaged over the pump-probe delay)  magnitude of $\left\langle\bm{v}_{\Omega}^{L}\right\rangle$.

\section{4. Temporal Geometry in Photoexcitation} %as an alternative channel of PI-MOCD}
The Berry curvature can also be associated with the excitation of bound states. Importantly, the general definition Eq.\,\eqref{eq:curvature_def} describes both cases, even though the specific expressions are different. This surprising result demonstrates the importance of the laser pulse polarization geometry: applying a circularly polarized pulse either to photoionize or to excite the molecule, one can address these two different "faces" of the Berry curvature.  
%Question from Ling: What does this sentence mean precisely?  That the notion of "curvature" can also be applied to the excitations that correspond to bound states, or that the previously defined curvature also applies to the excitations that correspond to bound states?  The former implies a different definition and expression concretely, and the latter implies the same definition and same expression.

Let a circularly polarized pump pulse excite a superposition of states in a chiral molecule. We now show that the subsequent photoionization by a linearly polarized probe pulse will lead to enantio-sensitive orientation of ions (PI-MOCD). Exchanging $\bm{E}_{{ki}}$ and $\bm{\mathcal{E}}_{i}$ in Eq.\,\eqref{eq:a_k}, using Eq.\,\eqref{eq:curvature_def} and following the same steps as in the previous section, we obtain the analog of Eq.\,\eqref{eq:_L_continuum}:
\begin{align}
\bm{\Omega}_{\mathrm{exc}}^{M}(k,\rho)%=i\langle\bm{\nabla}_{\bm{E}}\psi|\times|\bm{\nabla}_{\bm{E}}\psi\rangle & 
=-|E_{\omega_{1}}||E_{\omega_{2}}|\mathrm{Im}\left\{\int\mathrm{d}\Theta_{k}\left(\bm{D}_{1}\cdot\bm{\mathcal{E}}_{{\bm{k}1}}\right)^{*}\left(\bm{D}_{2}\cdot\bm{\mathcal{E}}_{{\bm{k}2}}\right) e^{i\omega_{12}\tau}\right\}\left( \bm{d}_{1}^M\times\bm{d}_{2}^M\right).
\label{eq:_L_bound}
\end{align}
Here the subscript 'bound' emphasizes the nature of the  Berry curvature, which now relies on bound states. 
% I moved the following part below the Binnet Cauchy identity like in the previous section, so that the equation for the yield in terms of Phi_B is now for the isotropic part. 
% Analogously to Eq.\,(\ref{eq:flux}),  the  phase is given by the flux of the  curvature trough the sphere in molecular frame uniting the points defined by the two Euler angles,  and is proportional to the  total, averaged over all random molecular orientations $\rho$, %$(\int d\rho=\frac{1}{8\pi^2}\int_{0}^{\pi} \sin \theta d\theta \int_{0}^{2\pi} d\phi \int_{0}^{2\pi}d\alpha)$, 
% enantio-sensitive photoionization probability:
% {%\color{red}
% \begin{align}
% W^{CL}(k)&=\int W^{CL}(n,\rho)d\rho=\frac{\sigma}{4\pi^2}|\bm{E}_{{1}}\cdot \bm{E}_{{2}}|
%   \int \bm{\mathsf{\Omega}}^{\mathsf{M}}(k,\rho) \cdot d\bm{S}^{\mathsf{M}}=\frac{\sigma}{4\pi^2}|\bm{E}_{{1}}\cdot \bm{E}_{{2}}|\Phi_B.
%   \label{eq:flux2} 
% \end{align}
% }
\begin{figure}
\includegraphics[width=0.86\textwidth, angle=0]{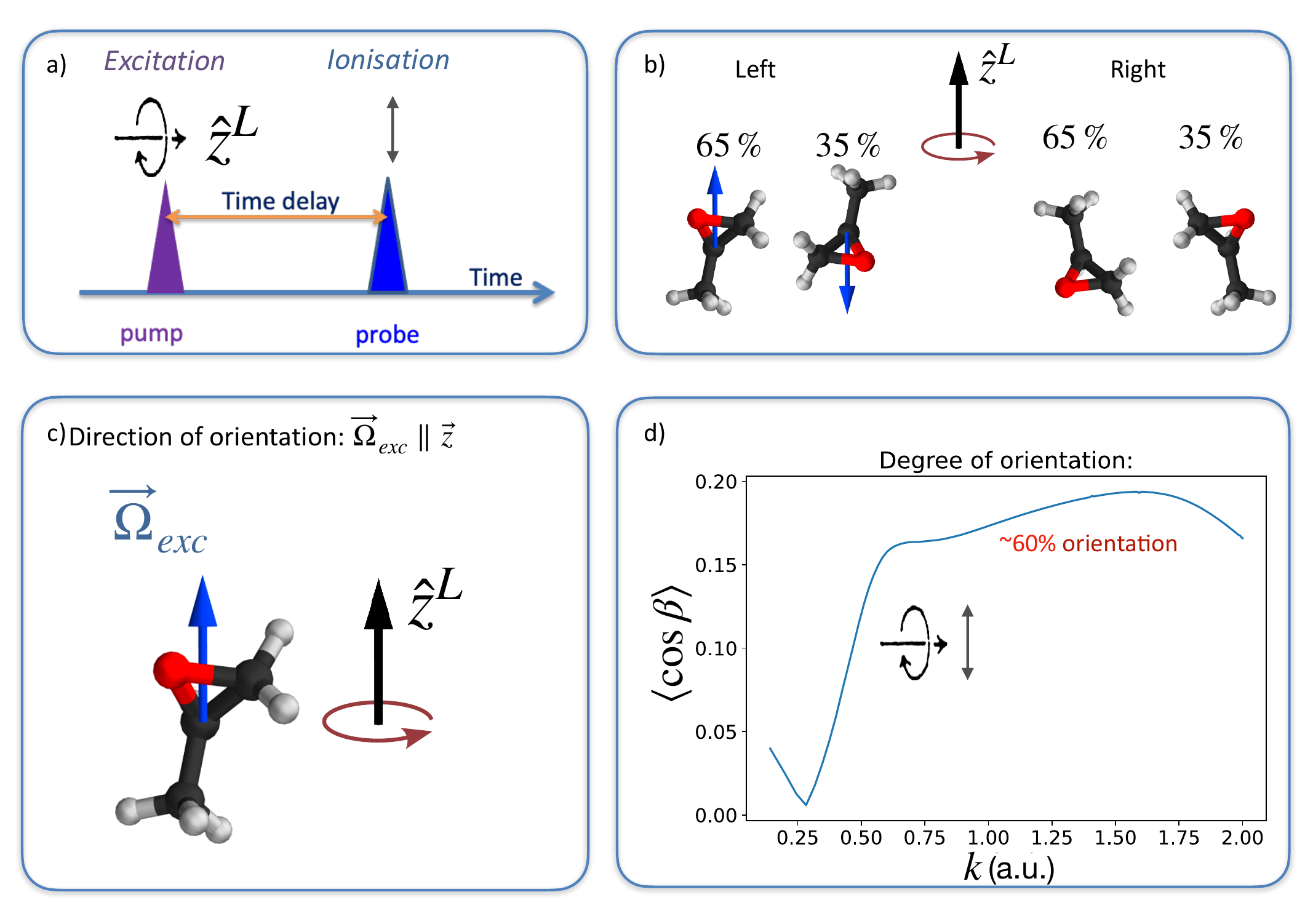}
\caption{Enantio-sensitive molecular orientation by excitation probed via photoionization. (a) A circularly polarized  pulse ($\hat{\bm{z}}^L$ is the direction of photon spin in the laboratory frame) excites a superposition of  LUMO and LUMO$_{+1}$ states in propylene oxide. (b) Molecules with $\bm{\Omega}_{exc}\parallel \hat{\bm{z}}^L$ are excited preferentially. A linearly polarized probe removes the excited electron, leading to enantio-sensitive delay-dependent orientation of cations. Estimated  percentages of "up" and "down" oriented species correspond to the maximal value of  $\langle \cos \beta\rangle\simeq 0.19$ and are obtained using the same model as in Ref.\,\cite{ordonez2023geometric} (c) Direction of molecular orientation is defined by the Berry curvature vector. (d) Expectation value  $\langle \cos \beta\rangle$ as a function of the photoelectron momentum $k$ (in atomic units). The values have been taken at the time-delay that maximizes $\left\langle\bm{v}_{\Omega}^{L}\right\rangle$  and have been normalized to the total yield averaged over the pump-probe time delay.}
\label{fig:exc}
\end{figure} 

%Using the Binet-Cauchy identity we obtain the isotropic component of the  curvature in bound states as
The orientationally averaged expression for the scalar product  $\bm{\Omega}\cdot\bm{v}_{\Omega}$, where $\bm{v}_{\Omega}$ is oriented along the direction of $\left( \bm{d}_{1}\times\bm{d}_{2}\right)$ in the molecular frame, can be derived in the same way as in the previous section:
\begin{align}
\langle\bm{{{\Omega}}}_{\mathrm{exc}}(k)\cdot\bm{v}_{\Omega}\rangle=-\frac{1}{3}C
\mathrm{Re}\left\{\int\mathrm{d}\Theta_{k}\bm{D}_{1}^{*}\cdot\bm{D}_{2}\right\}
%|E_{\omega_{k1}}||E_{\omega_{k2}}|\left(\bm{\mathcal{E}}_{{\bm{k}1}}^{*}\cdot \bm{\mathcal{E}}_{{\bm{k}2}}\right)
%+\emph{Im}\left\{\int\mathrm{d}\Theta_{k}\bm{D}_{1}\cdot\bm{D}_{2}\right\}\sin(\omega_{12}\tau)\right)
\left( \bm{d}_{1}\times\bm{d}_{2}\right)\cdot\bm{v}_{\Omega}\sin(\omega_{12}\tau),
\label{eq:MIS__bound}
\end{align}
where $C\equiv|\mathcal{E}_{\omega_{\bm{k}1}}||{E}_{\omega_{1}}||\mathcal{E}_{\omega_{\bm{k}2}}||{E}_{\omega_{2}}|$ and  we took into account that for the time-reversal-
invariant excited states $\psi_1$ and $\psi_2$, the time-reversal symmetry implies that (see Appendix C):
\begin{equation}
\mathrm{Im}\left\{\int\mathrm{d}\Theta_{k}\bm{D}_{1}^{*}\cdot\bm{D}_{2}\right\} = 0.
\label{auxillary}
\end{equation} 
Note that $\left( \bm{d}_{1}\times\bm{d}_{2}\right)\cdot\bm{v}_{\Omega}=\upsilon \left| \bm{d}_{1}\times\bm{d}_{2}\right|\bm{v}_{\Omega}$, meaning that Eq.\,\eqref{eq:MIS__bound} can also be rewritten as
\begin{align}
\langle\bm{{{\Omega}}}_{\mathrm{exc}}(k)\cdot\bm{v}_{\Omega}\rangle=-\frac{1}{3}\upsilon C
\mathrm{Re}\left\{\int\mathrm{d}\Theta_{k}\bm{D}_{1}^{*}\cdot\bm{D}_{2}\right\}
%|E_{\omega_{k1}}||E_{\omega_{k2}}|\left(\bm{\mathcal{E}}_{{\bm{k}1}}^{*}\cdot \bm{\mathcal{E}}_{{\bm{k}2}}\right)
%+\emph{Im}\left\{\int\mathrm{d}\Theta_{k}\bm{D}_{1}\cdot\bm{D}_{2}\right\}\sin(\omega_{12}\tau)\right)
\left| \bm{d}_{1}\times\bm{d}_{2}\right|\sin(\omega_{12}\tau).
\label{eq:MIS__bound2}
\end{align}

%The enantio-sensitive molecular orientation by ionization (PI-MOCD) for the linear pump - circular probe excitation is associated with the preferential photoexcitation of the molecules with the curvature  oriented along the photon spin. The linearly polarized probe simply removes the excited electron.
Using Eq.\,\eqref{eq:observables_intro} together with Eq.\,\eqref{eq:_L_bound} we can calculate (see Appendix D) the orientation averaged value of 
 \begin{equation} \left\langle\bm{v}_{\Omega}^{L}\right\rangle^{\mathrm{\circlearrowleft,\updownarrow}}=N\sigma \left\langle\left(\bm \Omega_{\mathrm{exc}}^L\cdot\hat{\bm{z}}^{L}\right)\bm{v}_{\Omega}^{L}\right\rangle=N\sigma\mathrm{R}^{(2,\mathrm{\circlearrowleft,\updownarrow})}\langle\bm \Omega_{\mathrm{exc}}\cdot\bm{v}_{\Omega}\rangle\hat{\bm{z}}^{L},
\label{eq:observables_bound}
\end{equation}
where $N^{-1}=\int_0^{2\pi} d\phi_{\tau}\int\mathrm{d}\rho\,W^{\mathrm{\circlearrowleft,\updownarrow}}(\rho,\tau)$ is a normalization factor, $\int_0^{2\pi} d\phi_{\tau}$ is averaging over the cycle of pump-probe delay, $W^{\mathrm{\circlearrowleft,\updownarrow}}(\rho,\tau)$ is ionization rate for a given orientation $\rho$, and
\begin{equation}
R^{(2,\mathrm{\circlearrowleft,\updownarrow})}\equiv\frac{2}{5}\left[1 - \frac{1}{2}\frac{Re\left\{\int d \Theta_k(\bm{D}^*_{1}\cdot\bm{v}_{\Omega})(\bm{D}_{2}\cdot\bm{v}_{\Omega})\right\}}{Re\left\{\int d \Theta_k(\bm{D}^*_{1}\cdot\bm{D}_{2})\right\}}\right].
\label{eq:orientation_full_factor_new}	
 \end{equation} 
 Evaluating it explicitly 
\begin{align}
\langle\bm{v}_{\Omega}^{\mathrm{L}}\rangle^{\mathrm{\circlearrowleft,\updownarrow}} & =N\int\mathrm{d}\rho\,W^{\mathrm{\circlearrowleft,\updownarrow}}(k,\rho)\bm{v}^{\mathrm{L}}_{\Omega}, %=\frac{1}{2}\sigma \int\mathrm{d}\varrho\,\left(\bm{\Omega}^{\mathrm{L}}_{\mathrm{ion}}(k,\rho)\cdot \hat{\bm{z}}^{\mathrm{L}}\right)\bm{v}^{\mathrm{L}}(\rho).
\label{eq:v_lin_circ_Lab}
\end{align}
we get the expression equivalent to Eq.\,\eqref{eq:observables_bound},   up to rearrangement of terms:
\begin{equation}
\left\langle\bm{v}_{\Omega}^{L}\right\rangle^{\mathrm{\circlearrowleft,\updownarrow}} =\frac{i\sigma}{60}CN\int\mathrm{d}\Theta_{k}\left[4\left(\bm{D}_{2}^{*}\cdot\bm{D}_{1}\right)\bm{v}_{\Omega}^{M}-\left(\bm{D}_{2}^{*}\cdot\bm{v}_{\Omega}\right)\bm{D}_{1}^M-\left(\bm{D}_{1}\cdot\bm{v}_{\Omega}\right)\bm{D}_{2}^{*M}\right]\left(\bm{d}_{2}^M\times\bm{d}_{1}^M\right)e^{i\omega_{21}\tau}\hat{\bm{z}}+\mathrm{c.c.}\label{eq:circ-lin_exact}
\end{equation}

Eq.\,(\ref{eq:observables_bound}) together with Eq.\,(\ref{eq:MIS__bound})  show that  PI-MOCD can also emerge due to the Berry curvature in bound states: molecules, in which the Berry curvature is  oriented along the spin of the laser light, are excited preferentially.

Thus, photoionization following photoexcitation creates enantio-sensitive orientation of both neutrals and cations. The expectation value of $\cos\beta$, where $\beta$ is the angle between the unit vector along the Berry curvature $\bm{v}_{\Omega}^{L}$ and the direction of photon spin,  is given by the normalized (to the total photoionization yield averaged over the pump-probe delay) magnitude of $\left\langle\bm{v}_{\Omega}^{L}\right\rangle$.
The orbital antisymmetric Berry curvature describes the orientation of cations in the dominant isotropic case. In the rare cases when the first, isotropic, term in Eq.\eqref{eq:circ-lin_exact} vanishes the last two terms originating from the alignment of molecular ensemble define the direction of enantio-sensitive orientation.%, which in this case is controlled by the direction of photionization dipoles $\bm{D}_{2}^{M}$ and $\bm{D}_{1}^{M}$. 

\section{5. Interplay of bound and continuum temporal geometries}

\begin{figure}
\includegraphics[width=0.86\textwidth, angle=0]{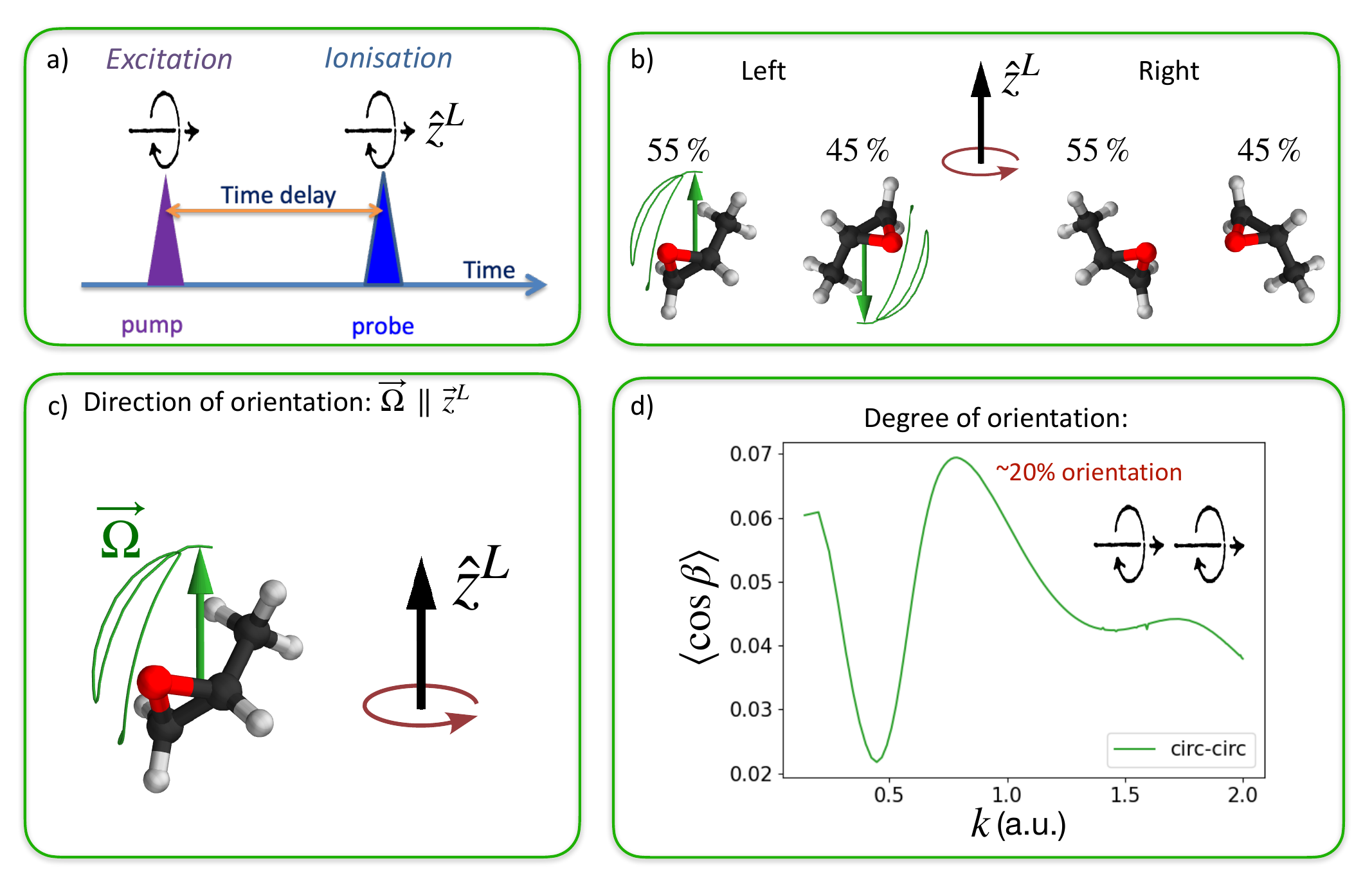}
\caption{Enantio-sensitive molecular orientation by excitation probed via photoionization. (a) A circularly polarized  pulse ($\hat{\bm{z}}^L$ is the direction of photon spin in the laboratory frame) excites a superposition of  LUMO and LUMO$_{+1}$ states in propylene oxide. (b) A circularly polarized probe removes the excited electron, leading to enantio-sensitive delay-dependent orientation of cations. Molecules with $\bm{\Omega}\parallel \hat{\bm{z}}^L$ are excited and ionised preferentially.
%As a function of photoelectron energy, the tip of the curvature vector traces the green curve in space. This green curve illustrates the sequence of orientations achieved for different final photoelectron momenta from 0.25 to 2 a.u.
The green curve represents the trajectory traced by the tip of the Berry curvature vector in space as a function of photoelectron energy, from 0.25 to 2 a.u. It shows how different photoelectron momenta lead to different orientation of the cation.
Estimated  percentages of "up" and "down" oriented species correspond to the maximal value of  $\langle \cos \beta\rangle\simeq 0.07$ and are obtained using the same model as in Ref.\,\cite{ordonez2023geometric} (c) Direction of molecular orientation is defined by the Berry curvature vector. (d)  Expectation value  $\langle \cos \beta\rangle$ as a function of the photoelectron momentum $k$ (in atomic units). The values have been taken at the time-delay that maximizes $\left\langle\bm{v}_{\Omega}^{L}\right\rangle$  and have been normalized to the total yield averaged over the pump-probe time delay. }
\label{fig:circ}
\end{figure} 

In the previous two sections, we considered photoexcitation followed by photoionization induced by pairs of pump-probe pulses with linear - circular or circular-linear polarizations correspondingly. In the first case, the Berry curvature has emerged due to the temporal geometry in photoelectron continuum states. In the second case, it has emerged due to the temporal geometry in bound states. What happens if both the pump and the probe pulses are circularly polarized? 

To answer this question, we need to apply Eqs.\,(\ref{eq:curvature_def})  to equations (\ref{eq:psi}-\ref{eq:a_k}), where the Fourier components of the linearly polarized fields $\bm{\mathcal{E}}_{2}$ and $\bm{\mathcal{E}}_{1}$ are substituted by the Fourier components of circularly polarized fields:  $\bm{\mathcal{E}}_{2}=>\bm{E}_{2}$ and
$\bm{\mathcal{E}}_{1}=>\bm{E}_{1}$. %Note that there is a difference  in using Eq.\,(\ref{eq:curvature_def}) in this case. 
%As detailed in Appendix B, the gradients $\bm{\nabla}_{\bm{e}}$ in Eq.\,(\ref{eq:curvature_def}) should be taken only with respect to fields carrying spin angular momentum, because only these fields uniquely define the orientation of the laboratory frame with respect to the molecular frame. 
%That is why in sections 3,4 the gradients $\bm{\nabla}_{\bm{e}}$  were  taken only with respect to the circularly polarised Fourier components of the field (i.e. probe field  in Section 3 and pump field in Section 4). 
Here, we have to evaluate the gradients in Eq.\,(\ref{eq:curvature_def}) with respect to both the pump field components and the probe field components. It leads to the following expression for the Berry curvature (see Appendix F):
\begin{align}
\label{eq:curv_L_mixed}
\bm{\Omega}^M(k,\rho)%=i\langle\bm{\nabla}_{\bm{E}}\psi|\times|\bm{\nabla}_{\bm{E}}\psi\rangle & 
&=\bm{\Omega}_{\mathrm{exc}}^M(k,\rho)+\bm{\Omega}^M_{\mathrm{ion}}(k,\rho)+\bm{\Omega}^M_{\mathrm{mix}}(k,\rho)
\end{align}
 where 
 \begin{align}
 \label{eq:_L_bound_circ}
\bm{\Omega}^M_{\mathrm{exc}}(k,\rho)%=i\langle\bm{\nabla}_{\bm{E}}\psi|\times|\bm{\nabla}_{\bm{E}}\psi\rangle & 
&=-2C\mathrm{Im}\left\{\int\mathrm{d}\Theta_{k}\left(\bm{D}_{1}\cdot\bm{e}\right)^{*}\left(\bm{D}_{2}\cdot\bm{e}\right) e^{i\omega_{12}\tau}\right\}\left( \bm{d}_{1}^M\times\bm{d}_{2}^M\right),\\
\bm{\Omega}^M_{\mathrm{ion}}(k,\rho)%&=i\langle\bm{\nabla}_{\bm{E}}\psi|\times|\bm{\nabla}_{\bm{E}}\psi\rangle \\\nonumber
&=-2C\mathrm{Im} \left\{\left(\bm{d}_{1}\cdot\bm{e}\right)^{*}\left(\bm{d}_{2}\cdot\bm{e}\right) e^{i\omega_{12}\tau}\int\mathrm{d}\Theta_{k}\bm{D}_{1}^{*M}\times\bm{D}_{2}^M\right\}.\\
\bm{\Omega}^M_{\mathrm{mix}}(k,\rho)=
&-2C\mathrm{Im}\left\{\int\mathrm{d}\Theta_{k}\left(\bm{d}_{1}\cdot\bm{e}\right)\left(\bm{D}_{2}\cdot\bm{e}\right)^{*} \left( \bm{D}_{1}^{*M}\times\bm{d}_{2}^M\right)e^{i\omega_{12}\tau}\right\}
\\\nonumber
&-2C\mathrm{Im}\left\{\int\mathrm{d}\Theta_{k}\left(\bm{D}_{1}\cdot\bm{e}\right)^{*}\left(\bm{d}_{2}\cdot\bm{e}\right) \left( \bm{d}_{1}^M\times\bm{D}_{2}^M\right)e^{i\omega_{12}\tau}\right\},
\label{eq:_L_cont_circ}
\end{align}
Here $C\equiv|E_{\omega_{\bm{k}1}}||E_{\omega_{1}}||E_{\omega_{\bm{k}2}}||E_{\omega_{2}}|$. Eq.\,(\ref{eq:curv_L_mixed}) combines bound and continuum Berry curvatures as detailed in  Eqs.\,(\ref{eq:_L_continuum}, \ref{eq:_L_bound}),  and also contains "mixed" terms, relying on both bound and continuum Berry curvatures.

The orientationally averaged scalar product of the Berry curvatures and a unit vector $\bm{v}_{\Omega}$ are:
\begin{align}
%\label{eq:_L_mixed}
\langle\bm{\Omega}_{\mathrm{exc}}(k,\rho)\cdot\bm{v}_{\Omega}\rangle%=i\langle\bm{\nabla}_{\bm{E}}\psi|\times|\bm{\nabla}_{\bm{E}}\psi\rangle & 
&=-\frac{1}{3}C \mathrm{Re}\left\{\int\mathrm{d}\Theta_{k}\left(\bm{D}_{1}^{*}\cdot\bm{D}_{2}\right) \right\}\left( \bm{d}_{1}\times\bm{d}_{2}\right)\cdot\bm{v}_{\Omega}\sin(\omega_{12}\tau),\\
\langle\bm{\Omega}_{\mathrm{ion}}(k,\rho)\cdot\bm{v}_{\Omega}\rangle%&=i\langle\bm{\nabla}_{\bm{E}}\psi|\times|\bm{\nabla}_{\bm{E}}\psi\rangle \\\nonumber
&=-\frac{1}{3}C\left(\bm{d}_{1}\cdot\bm{d}_{2}\right)\bm{\mathsf{P}}^{+}_{12}(k)\cdot\bm{v}_{\Omega}\sin(\omega_{12}\tau).\\
\langle\bm{\Omega}_{\mathrm{mix}}(k,\rho)\cdot\bm{v}_{\Omega}\rangle=
&-\frac{1}{3}C\mathrm{Im}\left\{\int\mathrm{d}\Theta_{k}\left(\bm{d}_{1}\cdot\bm{D}_{2}^{*}\right) \left( \bm{D}^*_{1}\times\bm{d}_{2}\right)\cdot\bm{v}_{\Omega}e^{i\omega_{12}\tau}\right\}
\\\nonumber
&-\frac{1}{3}C\mathrm{Im}\left\{\int\mathrm{d}\Theta_{k}\left(\bm{D}_{1}^{*}\cdot\bm{d}_{2}\right) \left( \bm{d}_{1}\times\bm{D}_{2}\right)\cdot\bm{v}_{\Omega}e^{i\omega_{12}\tau}\right\},
\end{align}
%\begin{equation} \left\langle\bm{v}_{\Omega}^{L}\right\rangle^{\mathrm{\circlearrowleft,\circlearrowleft}}=N\sigma \left\langle\left(\bm \Omega^L\cdot\hat{\bm{z}}^{L}\right)\bm{v}_{\Omega}^{L}\right\rangle
%label{eq:observables_mixed}
%\end{equation}
 Evaluating $\left\langle \bm{v}_{\Omega}^{\mathrm{L}}\right\rangle^{\mathrm{\circlearrowleft}}$ explicitly without involving the Berry curvature:
\begin{align}
\label{eq:circ-circ}
\left\langle \bm{v}_{\Omega}^{\mathrm{L}}\right\rangle^{\mathrm{\circlearrowleft}}  & =\frac{i\sigma}{30}CN\int\mathrm{d}\Theta_{k}\bigg\{\left(\bm{d}_{2}\cdot\bm{d}_{1}\right)\left[\bm{v}_{\Omega}\cdot\left(\bm{D}_{2}^{*}\times\bm{D}_{1}\right)\right]+\left(\bm{D}_{2}^{*}\cdot\bm{D}_{1}\right)\left[\bm{v}_{\Omega}\cdot\left(\bm{d}_{2}\times\bm{d}_{1}\right)\right]\nonumber \\
 & +\left(\bm{d}_{2}\cdot\bm{D}_{1}\right)\left[\bm{v}_{\Omega}\cdot\left(\bm{D}_{2}^{*}\times\bm{d}_{1}\right)\right]+\left(\bm{D}_{2}^{*}\cdot\bm{d}_{1}\right)\left[\bm{v}_{\Omega}\cdot\left(\bm{d}_{2}\times\bm{D}_{1}\right)\right]\bigg\}
 e^{i\omega_{21}\tau}\hat{\bm{z}}+\mathrm{c.c.}
\end{align}
and comparing it to$\left\langle\bm{v}_{\Omega}^{L}\right\rangle^{\mathrm{\circlearrowleft,\circlearrowleft}}$:
\begin{equation} \left\langle\bm{v}_{\Omega}^{L}\right\rangle^{\mathrm{\circlearrowleft,\circlearrowleft}}=N\sigma \left\langle\left(\bm \Omega^L\cdot\hat{\bm{z}}^{L}\right)\bm{v}_{\Omega}^{L}\right\rangle=N\sigma R^{(2,{\mathrm{\circlearrowleft,\circlearrowleft}})}\langle\bm \Omega\cdot\bm{v}_{\Omega}\rangle\hat{\bm{z}}^{L},
\label{eq:observables_mixed_R}
\end{equation}
where $\Omega$ is defined by  Eq.\eqref{eq:curv_L_mixed} we find that 
%$\langle\bm \Omega_{\mathrm{mix}}\cdot\bm{v}_{\Omega}\rangle$ we find that 
the factor $R^{(2,{\mathrm{\circlearrowleft,\circlearrowleft}})}=\frac{1}{10}$  is purely numerical due to preserved cylindrical symmetry.

%We shall use Eqs.\,(\ref{eq:lin-circ},\ref{eq:circ-lin},\ref{eq:circ-circ}) to compare the strength and direction of the PI-MOCD for all three polarization sequences, addressing different origins (continuum, bound and mixed) of temporal geometries. 
%Finally, one can verify that the expression for the dichroic yield $W_{\mathrm{\circlearrowleft,\circlearrowleft}}-W_{\mathrm{\circlearrowright,\circlearrowright}}$ via geometric phase also remains invariant, because the sum of the curvatures (and the sum of the geometric phases) only contains terms independent of $\sigma$, just like the curvature and the geometric phase in the one-photon case: 
%\begin{align}
%W_{\mathrm{\circlearrowleft,\circlearrowleft}}-W_{\mathrm{\circlearrowright,\circlearrowright}}&=\frac{\sigma}{4\pi}
%  \int \left(\bm{{\Omega}}^M_{\circlearrowleft,\circlearrowleft}(k,\theta,\phi)+\bm{{\Omega}}^M_{\circlearrowright,\circlearrowright}(k,\theta,\phi)\right) \cdot d\bm{S}^{\mathsf{M}}= \frac{\sigma}{4\pi}\left(\Phi_{\circlearrowleft,\circlearrowleft}+\Phi_{\circlearrowright,\circlearrowright}\right).
%  \label{eq:flux_circ_circ} 
%\end{align}

\section{6. Opportunities for control of geometric observables}
%Question from Ling: I wish I had clearly been given to understand whether this is a new finding, why it is new, and why it is important or a new way of controlling and studying chiral molecules.  Does the "switching" refer to the process of photoionization, i.e. turning the molecule in an excited, but still bound, state to the ionized state?   
PI-MOCD is the first geometric observable that is directly proportional to the orbital antisymmetric Berry curvature, relying on the geometry of  either  bound or  continuum states. Importantly, the the orbital antisymmetric Berry curvature is a dynamical property, as it relies on the excitation of currents and reflects their temporal geometry. Dynamics offers several opportunities for control. The direction of enantio-sensitive molecular orientation is defined by the direction of the  orbital antisymmetric Berry curvature vector in the molecular frame. The Berry curvature vectors relying on the vector product of bound or continuum dipoles naturally have  different directions: $\bm{\Omega}_{\mathrm{exc}}\parallel\bm{d}_1 \times \bm{d}_2$ and  $\bm{\Omega}_{\mathrm{ion}}\parallel\mathrm{Re} \{\int\mathrm{d}\Theta_{k}\bm{D}_{1}^{*}\times\bm{D}_{2}\}$. Moreover,  both orientations are state-specific (see Figs.\,\ref{fig:ion},\ref{fig:exc} (b) for propylene oxide), which presents several opportunities for controlling the enantio-sensitive direction of molecular orientation.

First, by changing the frequency of the pump pulse, one can address different bound states and thus control both the magnitude and direction of the Berry curvature. Second,  controlling the polarization of the pump-probe pulse sequence also allows us to control the magnitude and the direction of the enantio-sensitive molecular orientation by switching the Berry curvature between its form defined by the geometry of the excited bound states to the one defined by the geometry of the continuum states.  Figs.\,\ref{fig:ion} and \ref{fig:exc} illustrate how the direction of the Berry curvature and respective molecular orientation changes if one switches polarizations of pump and probe pulses from circular (linear) to linear (circular).  Third, by changing the frequency of the probe pulse in the case of linear-circular pulse sequence one can change the direction of the Berry curvature as shown in Fig.\,\ref{fig:ion}(b).   

\begin{figure}
\includegraphics[width=0.8\textwidth]{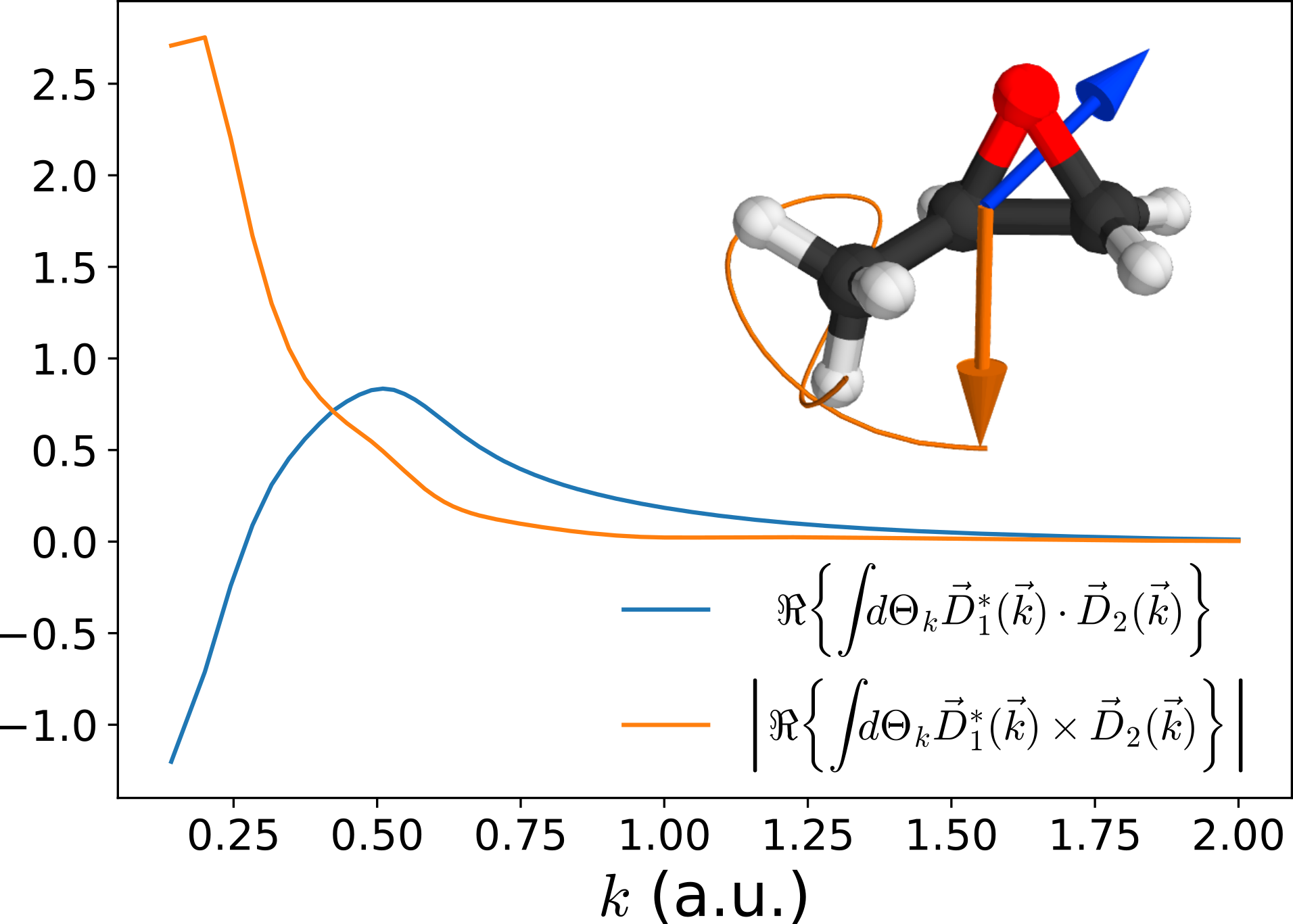}
\caption{Photoelectron momentum dependent terms $\mathrm{Re} \{\int\mathrm{d}\Theta_{k}\bm{D}_{1}^{*}\cdot\bm{D}_{2}\}$ and $|\mathrm{Re} \{\int\mathrm{d}\Theta_{k}\bm{D}_{1}^{*}\times\bm{D}_{2}\}|$, controlling the PI-MOCD in case of circular-linear pulse sequence (blue) and linear-circular pulse sequence (orange). The inset shows the molecule together with the molecular axis getting oriented in each case; $\bm{d}_1 \times \bm{d}_2$ in the circular-linear case (blue arrow), and $\mathrm{Re} \{\int\mathrm{d}\Theta_{k}\bm{D}_{1}^{*}\times\bm{D}_{2}\}$ in the linear-circular case (orange arrow). The latter changes as a function of photoelectron momentum $k$ and is shown for $k$=0.2 a.u. As $k$ increases, the tip of the orange arrow traces the orange curve.}
\label{fig:PI-MOCD-1}
\end{figure}

For a given set of intermediate bound states excited by a pump pulse and continuum states populated by the probe pulse,  the magnitudes of $\bm{{{\Omega}}}_{\mathrm{ion}}$ and  $\bm{{{\Omega}}}_{\mathrm{exc}}$   differ only in their dipole factors. The bound transition dipole contributions to the magnitudes of $\bm{{{\Omega}}}_{\mathrm{ion}}$ and  $\bm{{{\Omega}}}_{\mathrm{exc}}$ are $\bm{d_1}\cdot\bm{d_2}=-7.3\times 10^{-2}$ a.u. and $|\bm{d}_1 \times \bm{d}_2|=4.9\times 10^{-2}$ a.u., respectively. The continuum dipole contributions $\mathrm{Re} \{\int\mathrm{d}\Theta_{k}\bm{D}_{1}^{*}\cdot\bm{D}_{2}\}$ and $|\mathrm{Re} \{\int\mathrm{d}\Theta_{k}\bm{D}_{1}^{*}\times\bm{D}_{2}\}|$ change as a function of the photoelectron energy and are shown in Fig.\,\ref{fig:PI-MOCD-1}.
These two parameters determine the $k$-dependence of   expectation value of the degree of orientation  in Figs.\,\ref{fig:ion},\ref{fig:exc}(c).

\begin{comment} 
\begin{figure}
\includegraphics[width=0.8\textwidth]{orientation.png}
\caption{Molecular orientation circular dichroism in pump-probe photoionization. Expected value of the $z$ component of a molecular vector $\bm{v}$ as a function of photoelectron momentum $k$. The values have been taken at the time-delay that maximizes $\langle v_z^{\mathrm{L}}\rangle$ and have been normalized to the total yield averaged over the pump-probe time delay. For the circular-linear case $\bm{v}$ is a unitary vector parallel to $\bm{\Omega}_{\mathrm{bound,iso}}$. For the linear-circular case $\bm{v}$ is a unitary vector parallel to $\bm{\Omega}_{\mathrm{cont,iso}}$.} 
\label{fig:orientation}
\end{figure}
\end{comment}
\section{Outlook}
Geometric magnetism in chiral molecules appears to be a ubiquitous phenomenon, which can be controlled by tailoring light pulses and by addressing various molecular states. The linear-circular and circular-linear pump-probe sequences have been used experimentally to observe time-resolved photoelectron circular dichroism (TR-PECD)\cite{Comby2016JPCL,wanie2024capturing} and photoexcitation circular dichroism (PXCD)\cite{beaulieu_photoexcitation_2018} in chiral molecules via detection of angular distribution of photoelectrons. We show that in both cases, the detection of molecular fragments leads to a complementary geometric observable, directly proportional to the orbital antisymmetric Berry curvature related to currents in continuum or bound states. 

Our results show that the temporal evolution of angular distribution of fragments (Eqs.\,(\ref{eq:MIS__bound2},\ref{eq:observables_bound})) and angular distribution of photoelectrons produced after inducing  the photo-excitation circular dichroism    (circular-linear pump-probe sequence)\cite{beaulieu_photoexcitation_2018} and controlled by the term  $\int\mathrm{d}\Theta_{k}\bm{k}\left(\bm{D}_{1}^{*}\cdot\bm{D}_{2}\right)$ have fundamentally different properties.

The former is proportional to the current in the bound states $\propto\sin(\omega_{12}\tau)$ and thus vanishes at zero pump-probe delay $\tau=0$, while the latter also includes terms $\propto\cos(\omega_{12}\tau)$ and does not vanish at $\tau=0$.
This property establishes PI-MOCD as a direct probe of charge-directed reactivity: PI-MOCD is proportional to the  Berry curvature (either in bound or in continuum states) and as such relies on time-reversal symmetry breaking, which can be introduced by exciting a current prior to photoionization \cite{ordonez2023geometric}. The direction of the current at the moment of ionization defines the orientation of molecular cations and therefore the direction of molecular fragmentation, highlighting the ability of the ultrafast electron dynamics to control subsequent dynamics of the nuclei. %can be induced by linearly polarized pulses \cite{ordonez2023geometric}, or circularly polarized pulses.
%PI-MOCD can also be induced in strong fields, using multiphoton or tunneling ionization. In these
%Enantio-sensitive orientation of the fragments can be probed by fragmentation.
Simulations \cite{wanie2024capturing} show that enantio-sensitive fragment asymmetry resulting from excitation of electronic wave-packet in Rydberg states in methyl lactate molecule in linear-circular polarization sequence  reaches  20\%, confirming strong enantio-sensitivity of the Berry curvature-driven geometric observables.

The control over the directions of enantio-sensitive orientations of molecular ions suggests new opportunities. For example, in the linear-circular sequence of the pulses the tip of the Berry curvature vector traces the orange curve in space (Fig.\,\ref{fig:exc}(b)) as a function of photoelectron energy. This orange curve illustrates the sequence of orientations of the Berry curvature vector achieved for different final photoelectron momenta from 0.25 to 2 a.u. Since enantio-sensitive molecular orientation follows the orientation of the Berry curvature vector,   polarization controlled pulses at free electron lasers (FELs), such as chirped circularly polarized pulses, can be used to force molecular ions to draw the trajectory shown in Fig.\,\ref{fig:exc}(b) (see orange trajectory) within a single experiment. Alternately, the set of experiments in which the frequency of the probe circularly polarized light  is tuned within several tens of eV will produce a variety of well controlled molecular orientations. 

%Additional control opportunities emerge due to excitations of different superpositions of bound states, leading to different currents, which in turn  lead to different orientations of blue and orange vectors in Fig.\,2 and different orientations and fragmentation directions of the cations in space. Simulations \cite{wanie2024capturing} show that enantio-sensitive fragment asymmetry resulting from excitation of electronic wave-packet in Rydberg states in methyl lactate molecule in linear-circular polarization sequence  reaches  20\%, confirming strong enantio-sensitivity of curvature-driven geometric observables.  %as the frequency of the chirped FEL pulse changes within $13 eV$ to yield electrons with k from 0.25 $a.u$ to 0.75 $a.u$, corresponding to frequencies in the range from XXX 

%The geometric magnetism in bound states suggests new recipes for merging topological and enantio-sensitive  response  in chiral molecules. In particular, the flux of mixed curvature in the configurations space of laser field orientations emerging in  the excitation by circularly polarized pump and probe pulses is proportional to the enantio-sensitive and dichroic part of the photoionization yield (Eqs.\,\ref{eq:flux_general_intro}, \ref{eq:flux_circ_circ}).  In analogy to the  so-called Thouless charge pump, one would expect to realize a quantized rate of enantio-sensitive charge transfer via diabolic points in rotational states. 

\section{Acknowledgments}
We gratefully acknowledge many enlightening discussions with Prof. Misha Ivanov. We gratefully acknowledge Prof. Vladimir Chernyak for discussions and lectures on the topic.
%A.F.O. acknowledges grants supporting research at ICFO: Agencia Estatal de Investigación (“Severo Ochoa” Center of Excellence CEX2019-000910-S, National Plan FIDEUA PID2019-106901GB-I00/10.13039 / 501100011033, FPI), Fundació Privada Cellex, Fundació Mir-Puig, Generalitat de Catalunya (AGAUR Grant No. 2017 SGR 1341, CERCA program) and EU Horizon 2020 Marie Skłodowska-Curie grant agreement No 101029393, supporting his research on chirality. 
A.F.O. and D.A. acknowledge funding from the Royal Society (URF/R1/201333, URF/ERE/210358 and URF/ERE/231177). O.S., A.R. P.M.M. gratefully acknowledge ERC-2021-AdG project ULISSES, grant agreement No 101054696. Views and opinions expressed are however those of the authors only and do not necessarily reflect those of the European Union or the European Research Council. Neither the European Union nor the granting authority can be held responsible for them.
 
\section{Appendices}
\subsection{Appendix A: Closed lift for a state vector $\ket{\psi_{\rm el}(t)}\bra{\psi_{\rm el}(t)}$}
{The closed lift is a concept from fiber bundle theory that allows one to identify the geometric phase outside the approximation of an adiabatic evolution.}
%Indeed, during the  adiabatic evolution the wave-function does not acquire the dynamical phase, therefore if a phase is accumulated as a result of such evolution it must be a geometric phase. }

{In the Aharonov-Anandan fiber bundle theory, one introduces the base space in which all pure state vectors $\ket{\psi_{\rm el}(t)}\bra{\psi_{\rm el}(t)}$ live. By definition, the state vectors do not possess any phase. Thus, when the system evolves along the closed contour in the base space, the state vector remains the same after this cyclic evolution along the contour $\gamma$ : $\ket{\psi_{\rm el}(t)}\bra{\psi_{\rm el}(t)}=\ket{\psi_{\rm el}(t+T)}\bra{\psi_{\rm el}(t+T)}.$}

{Wave-functions corresponding to state vectors $\ket{\psi_{\rm el}(t)}\bra{\psi_{\rm el}(t)}$ at every point in base space $\ket{\psi_{\rm el}(t)}e^{i\phi}$ "live" on a fiber, which "grows vertically" from the base space while the phase $\phi$ is taking all possible values along 
this vertical direction (the fiber). These wave-functions are said to constitute "lifts" of the state vector $\ket{\psi_{\rm el}(t)}\bra{\psi_{\rm el}(t)}$  from the base space into the fiber space (at every point on the contour $\gamma$ in the base space). The wave-function that does not accumulate a phase as a result of cyclic evolution of a state vector on the contour $\gamma$ such that $\ket{\psi^{0}_{\rm el}(t)}=\ket{\psi^{0}_{\rm el}(t+T)}$ is called a closed lift.}

{We can show that the wave-function $\psi_{el}^{0}(t,\bm{e}(\rho),\bm{r})$ defined by Eq.\,\eqref{eq:TDSE_local} does not accumulate a phase as a result of cyclic evolution of rotational degrees of freedom and therefore presents the  closed lift of the state vector $\ket{\psi_{\rm el}(t)}\bra{\psi_{\rm el}(t)}$.}

{In our case, one cycle of evolution corresponds to one full rotation of the molecular frame $\psi_{el}^{0}(t,\bm{e}(\rho),\bm{r})$ from some initial  orientation characterised by Euler angles $\rho_0=\{\theta_0,\phi_0,\alpha_0\}$ with respect to the lab frame  to the exact  same orientation $\rho_0+2\pi=\{ \theta_0+2\pi,\alpha_0,\phi_0+2\pi\}$.
The wave-functions $\psi_{el}^{0}(t,\bm{e}(\rho_0),\bm{r})$ and $\psi_{el}^{0}(t,\bm{e}(\rho_0+2\pi),\bm{r})$ can be obtained from the same lab-frame wave-function using an appropriate sequence of unitary transformations, such as e.g. rotation around $\hat{n}$-axes by the angle $\xi$: $\psi_{el}^{0}(t,\bm{e}(\rho_0+2\pi),\bm{r})=e^{-i\xi(\hat{n}\cdot\hat{L})}\psi_{el}^{0}(t,\bm{r})=\psi_{el}^{0}(t,\bm{e}(\rho_0),\bm{r})$. Since these unitary transformations do not impart any relative phases on the transformed wave-functions $\psi_{el}^{0}(t,\bm{e}(\rho_0+2\pi),\bm{r})$ and 
$\psi_{el}^{0}(t,\bm{e}(\rho_0),\bm{r})$, and the same arguments remain valid for every point $\rho_0$, the wave-function $\psi_{el}^{0}(t,\bm{e}(\rho),\bm{r})$ presents a closed lift.}

\subsection{Appendix B: Derivation of Eq.\,\eqref{eq:S}}
Previously, we have derived Eqs.\,(\ref{eq:connection_t}, \ref{eq:curvature_def}) using the concept of adiabatic evolution of the rotational degree of freedom, introducing two different time scales for the evolution of electronic and rotational degrees of freedom. We now extend our derivation to a general case, including non-adiabatic evolution.
We follow the arguments of Ref.\,\cite{aharonov_anandan} and introduce a rotational Hamiltonian $H_{rot}(t)$, responsible for the cyclic non-adiabatic evolution of the rotational wave-function $\psi_{rot}(\rho,t)$: 
\begin{equation}
i\frac{d}{dt}\psi_{rot}(\rho,t)=H_{rot}(t)\psi_{rot}(\rho,t).
\label{eq:Hrot}
\end{equation}
For example, $H_{rot}(t)$ could encapsulate a sequence of microwave fields driving  resonant transitions in three rotational levels of a chiral molecule.
Since our focus is laser-induced coupling between electronic and rotational degrees of freedom, which arises due to the sensitivity of the electronic dynamics to the mutual orientation between the molecule and the laser field, we omit the standard Coriolis coupling between electronic and rotational degrees of freedom in the zero approximation.\\
Staring from the TDSE for the full wave-function including the electronic and rotational degrees of freedom:
\begin{equation}
i\frac{d}{dt}\Psi(\bm{r},\rho,t)=\bigg[H_{el}+\bm{r}\cdot\bm{E}(\rho,t)+H_{rot}(t)\bigg]\Psi(\bm{r},\rho,t)
\end{equation}
and using the ansatz:
\begin{equation}
\Psi(\bm{r},\rho,t)=\psi_{el}(\bm{r},\rho,t)\psi_{rot}(\rho,t),
\label{eq:anzatz}
\end{equation}
we first obtain
\begin{eqnarray}
i\psi_{rot}(\rho,t)\frac{d}{dt}\psi_{el}(\bm{r},\rho,t)+i\psi_{el}(\bm{r},\rho,t)\frac{d}{dt}\psi_{rot}(\rho,t)=\bigg[H_{el}+\bm{r}\cdot\bm{E}(\rho,t)+H_{rot}(t)\bigg]\psi_{el}(\bm{r},\rho,t)\psi_{rot}(\rho,t).
\label{eq:TDSE_1}
\end{eqnarray}
Multiplying the TDSE (Eq.\,\eqref{eq:TDSE_1}) by $\psi_{rot}^{*}(\rho,t)$ from the left and integrating over $\rho$ yields the following equation:
\begin{eqnarray}\label{eq:TDSE_2}
i\int d\rho|\psi_{rot}(\rho,t)|^2\frac{d}{dt}\psi_{el}(\bm{r},\rho,t)+i\int d\rho \psi_{el}(\bm{r},\rho,t)\psi_{rot}^{*}(\rho,t)\frac{d}{dt}\psi_{rot}(\rho,t)&=&\\\nonumber \int d\rho \psi_{rot}^{*}(\rho,t)\bigg[H_{el}+\bm{r}\cdot\bm{E}(\rho,t)\bigg]\psi_{el}(\bm{r},\rho,t)\psi_{rot}(\rho,t)\\\nonumber-i\int d\rho \frac{d}{dt}[\psi_{rot}^{*}(\rho,t)]\psi_{el}(\bm{r},\rho,t)\psi_{rot}(\rho,t),
\end{eqnarray}
where we have used Eq.\,\eqref{eq:Hrot} in the last term applying the rotational Hamiltonian $H_{rot}(t)$ to the bra-vector on the left. Note, that in doing so we have avoided the assumption of the adiabatic decoupling of the electronic and rotational degrees of freedom. Indeed, the adiabatic decoupling implies that $H_{rot}$ does not affect the electronic wave-function, 
while here it clearly does so via the last term in Eq.\,\eqref{eq:TDSE_2}. In the following we shall use the adiabatic decoupling to derive the equation for the closed lift solution (see Eqs.\,(\ref {eq:BO}-\ref{eq:TDSE_local_fin})).\\
Eq.\,\eqref{eq:TDSE_2} takes the  following form:
\begin{eqnarray}\label{eq:term_2}
i\int d\rho|\psi_{rot}(\rho,t)|^2\frac{d}{dt}\psi_{el}(\bm{r},\rho,t)+i\int d\rho \psi_{el}(\bm{r},\rho,t)\frac{d}{dt}|\psi_{rot}(\rho,t)|^2&=&\\\nonumber \int d\rho \psi_{rot}^{*}(\rho,t)\bigg[H_{el}+\bm{r}\cdot\bm{E}(\rho,t)\bigg]\psi_{el}(\bm{r},\rho,t)\psi_{rot}(\rho,t).
\end{eqnarray}
We shall now assume that the  rotational wave-packet is narrow. This assumption can be realized in an experiment by preparing a narrow rotational wavepacket with a pump pulse and using a short
probe pulse with time-dependent polarization, rotating with respect to the frame established
by the pump pulse.
Introducing $\dot{\overline{\rho}}(t)\equiv \int d\rho \frac{d}{dt}|\psi_{rot}(\rho,t)|^2\rho=\frac{d}{dt}\int d\rho |\psi_{rot}(\rho,t)|^2\rho=\frac{d}{dt}\overline{\rho}(t)$ and expanding $E_i(\rho,t)=E_i(\rho_0,t)+\nabla_{\rho_0}E_i(\rho_0,t)(\rho-\rho_0)$, and $\psi_{el}(\bm{r},\rho,t)$ up to the first order around $\overline{\rho}(t)$, we obtain for the integral $\int d\rho \psi_{el}(\bm{r},\rho,t)\frac{d}{dt}|\psi_{rot}(\rho,t)|^2$ the following expansion:
\begin{equation}
\int d\rho \psi_{el}(\bm{r},\rho,t)\frac{d}{dt}|\psi_{rot}(\rho,t)|^2= \int d\rho \psi_{el}(\bm{r},\overline{\rho}(t),t)\frac{d}{dt}|\psi_{rot}(\rho,t)|^2+\nabla_{\overline{\rho}}\psi_{el}(\bm{r},\overline{\rho}(t),t)\cdot\dot{\overline{\rho}}(t).
\label{eq:geom_phase}
\end{equation}
Here we took into account that $\int d\rho |\psi_{rot}(\rho,t)|^2=1$, which leads to $\frac{d}{dt}\int d\rho |\psi_{rot}(\rho,t)|^2=0$. Combining this term with the other terms, we obtain the following TDSE for the "global" wave-function $\psi_{el}(\bm{r},\overline{\rho}(t),t)$:
\begin{equation}
i\frac{\partial}{\partial t}\psi_{el}(\bm{r},\overline{\rho}(t),t)+i\nabla_{\overline{\rho}}\psi_{el}(\bm{r},\overline{\rho}(t),t)\cdot\dot{\overline{\rho}}(t)=\bigg[H_{el}+\bm{r}\cdot\bm{E}(\overline{\rho}(t),t)\bigg]\psi_{el}(\bm{r},\overline{\rho}(t),t).
\label{eq:TDSE_global1}
\end{equation}
Note that $\overline{\rho}(t)=\int d\rho \rho |\psi_{rot}(\rho,t)|^2$ follows the evolution of the center of the rotational wave-packet and therefore is controlled by $H_{rot}(t)$.\\
Let's derive  an approximate solution $\psi_{el}^{0}(\bm{r},\rho,t)$ for the case when the rotational Hamiltonian itself does not affect the electronic dynamics, i.e.  
\begin{equation}
H_{rot}(t)\psi_{el}^{0}(\bm{r},\rho,t)\psi_{rot}(\rho,t)=\psi_{el}^{0}(\bm{r},\rho,t)H_{rot}(t)\psi_{rot}(\rho,t).
\label{eq:BO}
\end{equation}
Applying this condition to  Eq.\,\eqref{eq:TDSE_1} yields
\begin{equation}
i\psi_{rot}(\rho,t)\frac{d}{dt}\psi_{el}^{0}(\bm{r},\rho,t)=\bigg[H_{el}+\bm{r}\cdot\bm{E}(\rho,t)\bigg]\psi_{el}^{0}(\bm{r},\rho,t)\psi_{rot}(\rho,t),
\label{eq:TDSE_local}
\end{equation}
demonstrating that $\psi_{el}^{0}(\bm{r},\rho,t)$ solves the "local" TDSE for fixed $\rho$:
\begin{equation}
i\frac{d}{dt}\psi_{el}^{0}(\bm{r},\rho,t)=\bigg[H_{el}+\bm{r}\cdot\bm{E}(\rho,t)\bigg]\psi_{el}^{0}(\bm{r},\rho,t).
\label{eq:TDSE_local_fin}
\end{equation}
Comparing Eq.\,\eqref{eq:TDSE_global1} with Eq.\,\eqref{eq:TDSE_local}  we find  that the "global" and the "local" solutions are related as follows:
\begin{equation}
\psi_{el}(\bm{r},\overline{\rho}(t),t)= e^{iS(\overline{\rho}(t))}\psi_{el}^{0}(\bm{r},\overline{\rho}(t),t).
\label{eq:geom_phase_direct}
\end{equation}
Substituting Eq.\,\eqref{eq:geom_phase_direct} to Eq.\,\eqref{eq:TDSE_global1}, we find:
\begin{eqnarray}
S(\overline{\rho}(t))=i\int_{\overline{\rho}(t_0)}^{\overline{\rho}(t)} d\rho' \langle\psi_{el}^{0}(\bm{r},\rho',t)|\nabla_{{\rho'}}\psi_{el}^{0}(\bm{r},\rho',t)\rangle,
\label{eq:geom_phase_der_2}
\end{eqnarray}
%The respective connection is $\bm{A}=i\langle\psi_{el}^{0}(\bm{r},\rho,t)|\nabla_{{\rho}}\psi_{el}^{0}(\bm{r},\rho,t)\rangle$.

Note that, by the virtue of equations Eq.\,(\ref{eq:BO}) and Eq.\,(\ref{eq:TDSE_local}), the only reason why $\psi_{el}^{0}(\bm{r},\rho,t)$ depends on $\rho$ is because the laser field in the molecular frame depends on $\rho$, i.e. $\psi_{el}^{0}(\bm{r},\rho,t)\equiv\psi_{el}^{0}(\bm{r},\bm{e}(\rho),t)$.

Indeed, $\psi_{el}^{0}(\bm{r},\rho,t)$ satisfies the TDSE for the electronic degrees of freedom
Eq.\ref{eq:TDSE_local_fin}. This equation, by definition, has no information about the rotational degree of freedom.  Thus, the gradient of  $\psi_{el}^{0}(\bm{r},\rho,t)$ with respect to $\rho$ can be rewritten in an equivalent form,  $\nabla_{{\rho}}\psi_{el}^{0}(\bm{r},\rho,t)=\nabla_{{\bm{e}}}\psi_{el}^{0}(\bm{r},\rho,t)\frac{\partial \bm{e}}{\partial \rho}$, as long as the laser field polarization $\bm{e}$ can be used to unambiguously define the laboratory frame with respect to which the molecular orientation $\rho$ is expressed. For example, circularly (elliptically) a polarized field polarization vector $\bm{e}$ satisfies this requirement. Hence the derivative should be taken with respect to $\bm{e}$. This leads  to the following equivalent expression for the geometric phase: 
\begin{eqnarray}
S(\overline{\rho}(t))=i\int_{\bm{e}(t_0)}^{\bm{e}(t)} \langle\psi_{el}^{0}(\bm{r},\rho',t)|\nabla_{\bm{e}}\psi_{el}^{0}(\bm{r},\rho',t)\rangle d\bm{e}\equiv \int_{\bm{e}(t_0)}^{\bm{e}(t)} \bm{A}(\bm{e})\cdot d\bm{e},
\label{eq:geom_phase_der_2}
\end{eqnarray}
where $\bm{A}(\bm{e})=\langle\psi_{el}^{0}(\bm{r},\bm{e},t)|\nabla_{\bm{e}}\psi_{el}^{0}(\bm{r},\bm{e},t)\rangle$.

\subsection{Appendix C: Time-reversal symmetry} 
To prove Eq.\,\eqref{auxillary} we extend the approach established in \cite{ordonez2023geometric} and introduce an operator $\hat{G}$:
\begin{align}
\int\mathrm{d}\Theta_{k}\bm{D}_{1}^{*}(\bm{k})\cdot\bm{D}_{2}(\bm{k})	& =\int\mathrm{d}\Theta_{k}\langle\psi_{\bm{k}}|\bm{r}|\psi_{1}\rangle^{*}\cdot\langle\psi_{\bm{k}}|\bm{r}|\psi_{2}\rangle \\
	& =\int\mathrm{d}\Theta_{k}\langle\psi_{1}|\bm{r}|\psi_{\bm{k}}\rangle\cdot\langle\psi_{\bm{k}}|\bm{r}|\psi_{2}\rangle \\
	& =\int\mathrm{d}\Theta_{k}\langle\psi_{1}|r_{i}P_{\bm{k}}r_{i}|\psi_{2}\rangle \\
	& =\langle\psi_{1}|r_{i}P_{k}r_{i}|\psi_{2}\rangle \\
	& =\langle\psi_{1}|\hat{G}|\psi_{2}\rangle 
\end{align}
We can now test the properties of this operator with respect to time-reversal symmetry.
\begin{equation}
\hat{G}\equiv\hat{r}_{i}\hat{P}_{k}\hat{r}_{i}
\end{equation}

\begin{equation}
\hat{G}^{\dagger}	=\hat{r}_{i}^{\dagger}\hat{P}_{k}^{\dagger}\hat{r}_{i}^{\dagger}
	=\hat{r}_{i}\hat{P}_{k}\hat{r}_{i}
	=\hat{G}
\end{equation}

\begin{equation}
\left[\hat{P}_{k},\hat{T}\right]=\left[\hat{r}_{i},\hat{T}\right]=0\Rightarrow\left[\hat{G},\hat{T}\right]=0
\end{equation}

\begin{equation}
\langle\alpha|\hat{G}|\beta\rangle=\langle\tilde{\alpha}|\hat{G}|\tilde{\beta}\rangle^{*}
\end{equation}

For $|\alpha\rangle=|\tilde{\alpha}\rangle$ and $|\beta\rangle=|\tilde{\beta}\rangle$, we get $\langle\alpha|\hat{G}|\beta\rangle=\langle\alpha|\hat{G}|\beta\rangle^{*}$, thus $\langle\alpha|\hat{G}|\beta\rangle\in\mathbb{R}$. This results yields Eq.\,\eqref{auxillary}. 

\subsection{Appendix D: Connection between the Coriolis Coupling and Temporal Geometry}

In this work we focus on the new, light-driven and light-controlled, mechanism of coupling between the electronic and rotational degrees of freedom, which is complementary to well known Coriolis coupling between the electronic and rotational degrees of freedom. Formally the difference between the standard Coriolis coupling and light-driven coupling can be seen from the following formal expressions.
The electronic wave-functions in laboratory frame $|\Psi\rangle^L$ and in the molecular frame $|\Psi\rangle^M$ are related by the unitary transformation encoding the set of  Euler angles $R$, which characterizes their mutual orientation $U(R)$: 
\begin{equation}
|\Psi\rangle^L=U(R)exp^{i\int\hat{H}^Mdt}|\Psi\rangle^M
\end{equation}
Since $R$ can depend on time, we obtain:
\begin{equation}
^L\!\!\langle \Psi|\frac{d}{dt}|\Psi\rangle^L idt= ^M\!\!\!\langle \Psi|U^{+}\frac{dU}{dt}|\Psi\rangle^M idt+^M\!\!\!\langle \Psi|\frac{d}{dt}|\Psi\rangle^M idt.
\end{equation}
Here the first term describes the Coriolis coupling, the second term describes the coupling due to temporal geometry: the dependence on the Euler angles comes from the laser field orientation, which appears in the argument of the electronic wave-function in the molecular frame because it also depends on Euler angles.

\subsection{Appendix E: Orientational averaging, derivation of Eq.\,\eqref{eq:observables} and Eq.\,\eqref{eq:_L_continuum_av} }
We shall use the following expressions:
\begin{align} 
    \int d\varrho l_{i\alpha} l_{j\beta} l_{k \gamma} l_{l \delta} 
    &= 
    \frac{1}{30} 
    \left( 
    \begin{array}{ccc}
        \kd{ij} \kd{kl} & \kd{ik} \kd{jl} & 
        \kd{il} \kd{jk} 
    \end{array} 
    \right) 
    \left( 
    \begin{array}{ccc}
        4 & -1 & -1 \\ 
        -1 & 4 & -1 \\ 
        -1 & -1 & 4 
    \end{array} 
    \right) 
    \left( 
    \begin{array}{c} 
         \kd{\alpha\beta} \kd{\gamma \delta}  \\
         \kd{\alpha \gamma} \kd{\beta \delta} \\ 
         \kd{\alpha \delta} \kd{\beta \gamma} 
    \end{array} 
    \right) \\ 
    &= 
    \frac{1}{30} 
    \Big[ 
    4\kd{ij} \kd{kl}\kd{\alpha\beta} \kd{\gamma \delta} -\kd{ik} \kd{jl}\kd{\alpha\beta} \kd{\gamma \delta} -\kd{il} \kd{jk}\kd{\alpha\beta} \kd{\gamma \delta} \\ 
    &-\kd{ij} \kd{kl}\kd{\alpha \gamma} \kd{\beta \delta} +4\kd{ik} \kd{jl}\kd{\alpha \gamma} \kd{\beta \delta} -\kd{il} \kd{jk}\kd{\alpha \gamma} \kd{\beta \delta} \\ 
    &-\kd{ij} \kd{kl}\kd{\alpha \delta} \kd{\beta \gamma} -\kd{ik} \kd{jl}\kd{\alpha \delta} \kd{\beta \gamma} +4\kd{il} \kd{jk}\kd{\alpha \delta} \kd{\beta \gamma} 
    \Big], 
\end{align} 

i.e., 
\begin{align} 
\label{Eq: Or_Av_4} 
    &\int d\rho 
    (\bm{a}^L \cdot \bm{u}^L) (\bm{b}^L \cdot \bm{v}^L) 
    (\bm{w}^L \cdot \bm{c}^L) \bm{x}^L \\\nonumber 
    &= 
    \frac{1}{30} 
    \Big[ 
    4(\bm{a}^L\cdot \bm{b}^L)\bm{c}^L (\bm{u}^M\cdot\bm{v}^M)(\bm{w}^M\cdot\bm{x}^M) \\ \nonumber 
    &-(\bm{a}^L\cdot \bm{c}^L)\bm{b}^L (\bm{u}^M\cdot\bm{v}^M)(\bm{w}^M\cdot\bm{x}^M) \\ \nonumber 
    &- (\bm{b}^L\cdot \bm{c}^L)\bm{a}^L (\bm{u}^M\cdot\bm{v}^M)(\bm{w}^M\cdot\bm{x}^M) \\ \nonumber 
    &-(\bm{a}^L\cdot \bm{b}^L)\bm{c}^L (\bm{u}^M\cdot\bm{w}^M)(\bm{v}^M\cdot\bm{x}^M) \\ \nonumber &+4(\bm{a}^L\cdot \bm{c}^L)\bm{b}^L (\bm{u}^M\cdot\bm{w}^M)(\bm{v}^M\cdot\bm{x}^M) \\ \nonumber 
    &- (\bm{b}^L\cdot \bm{c}^L)\bm{a}^L (\bm{u}^M\cdot\bm{w}^M)(\bm{v}^M\cdot\bm{x}^M) \\ \nonumber 
    &-(\bm{a}^L\cdot \bm{b}^L)\bm{c}^L (\bm{u}^M\cdot\bm{x}^M)(\bm{v}^M\cdot\bm{w}^M) \\ \nonumber 
    &-(\bm{a}^L\cdot \bm{c}^L)\bm{b}^L (\bm{u}^M\cdot\bm{x}^M)(\bm{v}^M\cdot\bm{w}^M) \\ \nonumber 
    &+4 (\bm{b}^L\cdot \bm{c}^L)\bm{a}^L (\bm{u}^M\cdot\bm{x}^M)(\bm{v}^M\cdot\bm{w}^M) 
    \Big]. 
\end{align}
We  derive Eq.\,\eqref{eq:observables}, which we list here for completeness: %Eq.\ref{eq:_L_continuum_av}.
\begin{equation} \left\langle\bm{v}_{\Omega}^{L}\right\rangle^{\mathrm{\updownarrow,\circlearrowleft}}=\sigma \left\langle\left(\bm \Omega_{\mathrm{ion}}\cdot\hat{\bm{z}}^{L}\right)\bm{v}_{\Omega}^{L}\right\rangle=\sigma\mathrm{R}\upsilon\langle\bm \Omega_{\mathrm{ion}}\cdot\bm{v}_{\Omega}\rangle\hat{\bm{z}}^{L},
\label{eq:observables_App}
\end{equation}
where
\begin{align}
\bm{\Omega}_{\mathrm{ion}}(k,\rho)%&=i\langle\bm{\nabla}_{\bm{E}}\psi|\times|\bm{\nabla}_{\bm{E}}\psi\rangle \\\nonumber
& =-|E_{\omega_{\bm{k}1}}||E_{\omega_{\bm{k}2}}|\left(\bm{d}_{1}\cdot\bm{\mathcal{E}}_{1}\right)^{*}\left(\bm{d}_{2}\cdot\bm{\mathcal{E}}_{2}\right)\bm{\mathsf{P}}^{+}_{12}(k)\sin(\omega_{12}\tau),
\label{eq:_L_continuum_App}
\end{align}
(Eq.\,\eqref{eq:_L_continuum}).
Substituting  Eq.\,\eqref{eq:_L_continuum_App} into Eq.\,\eqref{eq:observables_App}  we obtain:
\begin{eqnarray}
\left\langle\bm{v}_{\Omega}^{L}\right\rangle^{\mathrm{\updownarrow,\circlearrowleft}}=-|E_{\omega_{\bm{k}1}}||E_{\omega_{\bm{k}2}}|\sin(\omega_{12}\tau)\int d\varrho \left(\bm{d}_{1}^{L}\cdot\bm{\mathcal{E}}_{1}^{L}\right)^{*}\left(\bm{d}_{2}^{L}\cdot\bm{\mathcal{E}}^{L}_{2}\right)\left(\bm{\mathsf{P}}^{+L}_{12}(k)\cdot\hat{z^L}\right)\bm{v}_{\Omega}^{L}
    \label{eq:vector
_App}
\end{eqnarray}
Applying Eq.\,\eqref{Eq: Or_Av_4} to Eq.\,\eqref{eq:vector
_App} we obtain 
\begin{align}
\left\langle \bm{v}_{\Omega}\right\rangle^{\mathrm{\updownarrow,\circlearrowleft}}  & =\frac{\sigma}{30}C\left[4\left(\bm{d}_{2}\cdot\bm{d}_{1}\right)\bm{v}_{\Omega}^{\mathrm{M}}-\left(\bm{d}_{2}\cdot\bm{v}_{\Omega}^{\mathrm{M}}\right)\bm{d}_{1}-\left(\bm{d}_{1}\cdot\bm{v}_{\Omega}^{\mathrm{M}}\right)\bm{d}_{2}\right]\cdot\bm{\mathsf{P}}^{+}_{12}(k)\sin(\omega_{12}\tau)\hat{\bm{z}}^{L},\label{eq:lin-circ
_App}
\end{align}
which is up to notations $\bm{\mathsf{P}}^{+}_{12}(k)=\upsilon|\bm{\mathsf{P}}^{+}_{12}(k)|\bm{v}_{\Omega}$ is equivalent to Eq.\,\eqref{eq:observables}.

Starting from
\begin{align}
\bm{\Omega}_{\mathrm{ion}}(k,\rho)\cdot\bm{v}_{\Omega}^{\mathrm{M}}%&=i\langle\bm{\nabla}_{\bm{E}}\psi|\times|\bm{\nabla}_{\bm{E}}\psi\rangle \\\nonumber
& =-|E_{\omega_{\bm{k}1}}||E_{\omega_{\bm{k}2}}|\left(\bm{d}_{1}\cdot\bm{\mathcal{E}}_{1}\right)^{*}\left(\bm{d}_{2}\cdot\bm{\mathcal{E}}_{2}\right)\left(\bm{\mathsf{P}}^{+}_{12}(k)\cdot\bm{v}_{\Omega}^{\mathrm{M}}\right)\sin(\omega_{12}\tau),
\label{eq:_L_continuum_App_mod}
\end{align}
we shall now evaluate $\langle\bm{\Omega}_{\mathrm{ion}}\cdot\bm{v}_{\Omega}^{\mathrm{M}}\rangle$.  Note that we only need to average the part that depends on mutual orientations of the pump pulse and molecule:
\begin{eqnarray}
   \int d\varrho \left(\bm{d}_{1}\cdot\bm{\mathcal{E}}_{1}\right)^{*}\left(\bm{d}_{2}\cdot\bm{\mathcal{E}}_{2}\right)= \frac{1}{3}\left(\bm{d}_{1}\cdot\bm{d}_{2}\right)\left(\bm{\mathcal{E}}_{1}^*\cdot\bm{\mathcal{E}}_{2}\right)=\frac{1}{3}\left(\bm{d}_{1}\cdot\bm{d}_{2}\right)|{\mathcal{E}}_{\omega_1}||{\mathcal{E}}_{\omega_2}|, 
\end{eqnarray}
yielding Eq.\,\eqref{eq:_L_continuum_av}:
\begin{align}
\langle\bm{\Omega}_{\mathrm{ion}}(k)\cdot\bm{v}_{\Omega}\rangle%&=i\langle\bm{\nabla}_{\bm{E}}\psi|\times|\bm{\nabla}_{\bm{E}}\psi\rangle \\\nonumber
&=-\frac{1}{3}\sigma\upsilon C\left(\bm{d}_{1}\cdot\bm{d}_{2}\right)|\bm{\mathsf{P}}^{+}_{12}(k)|\sin(\omega_{12}\tau).
\label{eq:_L_continuum_av_App}
\end{align}

\subsection{Appendix F: Connection between the Berry Curvature and the Photoionization Yield}
In this section, we calculate the photoionization yield $W\equiv\int |a_{\bm{k}}|^2 d\Theta_k$ for two-photon processes considered  in this work. For example, consider a sequence of linear and circularly polarized pulses:
\begin{equation}
\bm{\mathcal{E}}^L_i={\mathcal{E}}_{\omega_i}\hat{\bm{x}}^L,\qquad\bm{E}^L_j=E_{\omega_{jk}}\frac{1}{\sqrt{2}}\left(\hat{\bm{x}}^L+i\sigma\hat{\bm{y}}^L\right)
\end{equation}
The photoionization amplitude $|a_{\bm{k}}|^2$ can be written as:
\begin{align}
  |a_{\bm{k}}|^2=\frac{1}{2}[|E_{\omega_{1k}}|^2\left|\left(\bm{d}^L_1\cdot\bm{\mathcal{E}}^L_1\right) \right|^2\left[\left|\bm{D}_1^L\cdot\hat{\bm{x}}^L\right|^2+\left|\bm{D}^L_1\cdot\hat{\bm{y}}^L\right|^2+i\sigma\left(\bm{D}_1^{*L}\times\bm{D}^L_1\right)\cdot \hat{\bm{z}}^L\right]\nonumber\\
  +\frac{1}{2}[|E_{\omega_{2k}}|^2\left|\left(\bm{d}^L_2\cdot\bm{\mathcal{E}}^L_2\right) \right|^2\left[\left|\bm{D}^L_2\cdot\hat{\bm{x}}^L\right|^2+\left|\bm{D}^L_2\cdot\hat{\bm{y}}^L\right|^2+i\sigma\left(\bm{D}_2^{*L}\times\bm{D}^L_2\right)\cdot \hat{\bm{z}}^L\right] \nonumber\\
  +\mathrm{Re}\left\{E_{\omega_{1k}}^*E_{\omega_{2k}}\left(\bm{d}^L_1\cdot\bm{\mathcal{E}}^L_1\right)^*\cdot\left(\bm{d}^L_2\cdot\bm{\mathcal{E}}^L_2\right) \left[\left(\bm{D}_1^{*L}\cdot\hat{\bm{x}}^L\right)\left(\bm{D}_2^L\cdot\hat{\bm{x}}^L\right)+\left(\bm{D}_1^{*L}\cdot\hat{\bm{y}}^L\right)\left(\bm{D}_2^L\cdot\hat{\bm{y}}^L\right)+i\sigma\left(\bm{D}_1^{*L}\times\bm{D}_2^L\right)\cdot \hat{\bm{z}}^L\right]\right\}
\end{align}
The dichroic (i.e. proportional to $\sigma$) photoionization yield is
\begin{eqnarray}
  \int|a_{\bm{k}}|^2 d\Theta_k=\sigma
  \mathrm{Re}\left\{E_{\omega_{1k}}^{*}E_{\omega_{2k}}\left(\bm{d}^L_1\cdot\bm{\mathcal{E}}^L\right)^*\cdot\left(\bm{d}^L_2\cdot\bm{\mathcal{E}}^L\right) \left[e^{i\omega_{12}\tau}\int i\left(\bm{D}_1^{*L}\times\bm{D}_2^L\right)d\Theta_k\right]\right\}\cdot \hat{\bm{z}}^L=\sigma\bm{\Omega}^L\cdot \hat{\bm{z}}^L,\\\nonumber
  %\equiv 2\sigma
  %\left(\bm{d}_1\cdot\bm{\mathcal{E}}\right)^*\cdot\left(\bm{d}_2\cdot\bm{\mathcal{E}}\right) \bm{\mathsf{P}}^{+}_{12}(k)\cdot \hat{\bm{z}}\sin\omega_{12}\tau=\Omega
\end{eqnarray}
where $\bm{\Omega}$ is given by Eq.\,\eqref{eq:_L_continuum}.\\

\subsection{Appendix G: Orbital Antisymmetric Berry Curvature for Circular Pump and Probe Fields: Derivation}
The wave function expressed in first order perturbation theory is
\begin{align}
    \ket{\psi} = \ket{0} + i \sum_{j} \bm{d}_j \cdot \bm{E}_j \ket{k} - \sum_j \int d\Theta_{k} (\bm{D}_j \cdot \bm{E}_{kj})(\bm{d}_j \cdot \bm{E}_j) e^{-i\omega_j\tau} \ket{\bm{k}}. 
\end{align}
The corresponding orbital antisymmetric Berry curvature is 
\begin{align}
    \bm{\Omega} = i \bra{\nabla_{\bm{e}} \psi} \times \ket{\nabla_{\bm{e}} \psi}
\end{align}

\begin{align}
    \ket{\nabla_{\bm{e}} \psi} = i \sum_{j} |E_{\omega_{j}}|\bm{d}_j \ket{k} - \sum_j |E_{\omega_{j}}||E_{\omega_{\bm{k}j}}|\int d\Theta_{k} \left[\bm{D}_j (\bm{d}_j \cdot \bm{e}) + (\bm{D}_j \cdot \bm{e})\bm{d}_j\right] e^{-i\omega_j\tau} \ket{\bm{k}}
\end{align}
Using the orthogonality of the basis, we get
\begin{align}
    -i\bm{\Omega} &= \sum_{j} |E_{\omega_{j}}|\bm{d}_j^* \times \bm{d}_j \nonumber\\
    &\quad + \sum_{jl} 
    |E_{\omega_{j}}||E_{\omega_{\bm{k}j}}||E_{\omega_{l}}||E_{\omega_{\bm{k}l}}| 
    \int d\Theta_{k}\left[\bm{D}_j^* (\bm{d}_j \cdot \bm{e})^* + (\bm{D}_j \cdot \bm{e})^*\bm{d}_j^*\right] \times \left[\bm{D}_l (\bm{d}_l \cdot \bm{e}) + (\bm{D}_l \cdot \bm{e})\bm{d}_l\right] e^{i\omega_{jl}\tau}
\end{align}
Since the dipole moments $\bm{d}_l$ are real the curvature simplifies to
\begin{align}
    -i\bm{\Omega} = \sum_{jl}|E_{\omega_{j}}||E_{\omega_{\bm{k}j}}||E_{\omega_{l}}||E_{\omega_{\bm{k}l}}|\int d\Theta_{k}\left[\bm{D}_j^* (\bm{d}_j \cdot \bm{e}^*) + (\bm{D}_j \cdot \bm{e})^*\bm{d}_j\right] \times \left[\bm{D}_l (\bm{d}_l \cdot \bm{e}) + (\bm{D}_l \cdot \bm{e})\bm{d}_l\right] e^{i\omega_{jl}\tau}
\end{align}
We employ the identity
\begin{align}
    \int d\Theta_{k}\bm{D}_j \times \bm{D}_j^* = 0\,,
\end{align}
such that terms satisfying $j = l$ vanish, 
\begin{align}
    -i\bm{\Omega} = \sum_{j\neq l}|E_{\omega_{j}}||E_{\omega_{\bm{k}j}}||E_{\omega_{l}}||E_{\omega_{\bm{k}l}}| 
    \int d\Theta_{k}\bigg[
    &(\bm{d}_j \cdot \bm{e}^*)(\bm{d}_l \cdot \bm{e})(\bm{D}_j^*\times\bm{D}_l)\nonumber\\
    +&(\bm{d}_j \cdot \bm{e}^*)(\bm{D}_l \cdot \bm{e})(\bm{D}_j^*\times\bm{d}_l)\nonumber\\
    +&(\bm{D}_j \cdot \bm{e})^*(\bm{d}_l \cdot \bm{e})(\bm{d}_j\times\bm{D}_l )\nonumber\\
    +&(\bm{D}_j \cdot \bm{e})^*(\bm{D}_l \cdot \bm{e})(\bm{d}_j\times\bm{d}_l )
    \bigg] e^{i\omega_{jl}\tau}. 
\end{align}
Due to the anti-symmetry of the cross product, the Berry curvature  can be expressed as 
\begin{align}
    \bm{\Omega} = -2C\int d\Theta_{k}\bigg[
    &\mathrm{Im}\left\{(\bm{d}_1 \cdot \bm{e}^*)(\bm{d}_2 \cdot \bm{e})(\bm{D}_1^*\times\bm{D}_2)e^{i\omega_{12}\tau}\right\}\nonumber\\
    +&\mathrm{Im}\left\{(\bm{d}_1 \cdot \bm{e}^*)(\bm{D}_2 \cdot \bm{e})(\bm{D}_1^*\times\bm{d}_2)e^{i\omega_{12}\tau}\right\}\nonumber\\
    +&\mathrm{Im}\left\{(\bm{D}_1 \cdot \bm{e})^*(\bm{d}_2 \cdot \bm{e})(\bm{d}_1\times\bm{D}_2 )e^{i\omega_{12}\tau}\right\}\nonumber\\
    +&\mathrm{Im}\left\{(\bm{D}_1 \cdot \bm{e})^*(\bm{D}_2 \cdot \bm{e})(\bm{d}_1\times\bm{d}_2 )e^{i\omega_{12}\tau}\right\}
    \bigg]\,,
\end{align}
where we defined 
\begin{align}
    C = |E_{\omega_1}||E_{\omega_{\bm{k}1}}||E_{\omega_2}||E_{\omega_{\bm{k}2}}|\,. 
\end{align} 

\subsection{Appendix H: Mathematical difference between the Aharonov-Anandan and Berry-Simons bundle}

%Let us consider a time dependent Hamiltonian $h(t) = h(R(t))$ whose time dependence is given by a path $R(t)$ of period $T$ in parameter space. This Hamiltonian can induce a closed path in the space of pure state density matrices such that 
%\begin{align} 
%    \Lambda(0) \equiv \ket{\psi(0)}\bra{\psi(0)} = \ket{\psi(T)}\bra{\psi(T)}\,.
%\end{align} 

Let us consider a Hamiltonian $h(\bm{R}(t))$ which induces a closed path of period $T$ in the space of pure state density matrices $\Lambda$ such that 
\begin{align} 
    \Lambda(0) \equiv \ket{\psi(0)}\bra{\psi(0)} = \ket{\psi(T)}\bra{\psi(T)}\,.
\end{align} 
The wavefunction (dynamical lift) corresponding to this path is quasi-periodic, i.e. is periodic up to a phasefactor which can be decomposed into a dynamical and a geometric contribution. In general, the dependence of the density matrix at time $t$ on the path $\bm{R}(t)$ is functional, which means that it depends on the whole history of $\bm{R}$ and not just on its value at time $t$. In this case, the underlying structure is the Aharonov-Anandan (AA) fiber bundle. The whole fiber space is the Hilbert space $\mathcal{H}$. The projection map
\begin{align}
    \pi : \ket{\psi} \to \ket{\psi}\bra{\psi}\,,
\end{align}
takes us to the base space of the AA bundle, the space of pure state density matrices. The Berry connection in the AA bundle is
\begin{align}
    \mathcal{A} = i \bra{\phi(t)} \frac{d}{dt} \ket{\phi(t)} dt\,,
\end{align}
where $\ket{\phi(t)}$ is the closed lift. It is defined s.t.
\begin{align}
    \ket{\phi(0)} = \ket{\phi(T)}
\end{align}
In some physical situations the dependency of the density matrix on $\bm{R}(t)$ is parametric, as e.g. in our case three Euler angles specifying a molecular rotation, governing the time evolution of the Hamiltonian. Above all, an adiabatic approximation with respect to the parameters $\bm{R}$ or a perturbative treatment of the wavefunction are to be listed here. Consequently, the $\bm{R}$-parameter space becomes isomorphic to the density matrix space and may, from a perspective of fiber bundle theory, be treated as the base space of the fiber bundle. In this case, the closed lifts are the instantaneous eigenstates of the Hamiltonian and the AA connection reduces to the Berry-Simons (BS) connection. This can be easily seen when writing down the expression for the geometric phase:
\begin{align}
    \gamma = \oint i \bra{\phi(t)} \frac{d}{dt} \ket{\phi(t)} dt \overset{\text{parametric dependence}}{=} \oint i \bra{\phi(\bm{R})} \nabla_{\bm{R}} \ket{\phi(\bm{R})} d\bm{R}\,.
\end{align}
This bundle is commonly referred to as BS bundle. Note that the formation of BS bundles solely depends on a parametric description of the underlying geometric problem and not on a cyclic evolution of the density matrix.

% takes us to the base space of the AA bundle, the space of pure state density matrices. In some physical situations the dependency of the density matrix can be traced back to some ($T$-periodic) parameter set $\bm{R}(t)$, as e.g. in our case three Euler angles specifying a molecular rotation, governing the time evolution of the Hamiltonian. In the process of solving the Schrödinger equation some approximations lead to a parametric dependence of the wavefunction on the geometric parameters $\ket{\psi(t)}\equiv \ket{\psi(\bm{R}(t))}$. Above all, an adiabatic approximation with respect to the parameters $\bm{R}$ or a perturbative treatment of the wavefunction are to be listed here. Consequently, the $\bm{R}$-parameter space becomes isomorphic to the density matrix space and may, from a perspective of fiber bundle theory, be treated as the base space of the fiber bundle. This bundle is commonly referred to as Berry-Simons bundle. 
% Note that the formation of Berry-Simons bundles solely depends on a parametric description of the underlying geometric problem and not on a cyclic evolution of the density matrix. 

\bibliography{Bibliography}
\end{document}